\documentclass[apj,numberedappendix]{emulateapj}
\usepackage{amssymb,amsmath}
\usepackage{graphicx}
\usepackage{appendix}
\usepackage{natbib}
\usepackage{hyperref}
\usepackage{enumerate}
\usepackage{multirow}
\usepackage{bm}
\usepackage{url}

\newcommand{\etal}{et~al.~}

\altaffiltext{\MIT}{Kavli Institute for Astrophysics and Space Research, Massachusetts Institute of Technology, 77 Massachusetts Avenue, Cambridge, MA 02139}
\altaffiltext{\Princeton}{Department of Astrophysical Sciences, Princeton University, 4 Ivy Lane, Princeton, NJ 08544-1001, USA}
\altaffiltext{\Waterloo}{Department of Physics \& Astronomy, University of Waterloo, Canada}
\altaffiltext{\CfA}{Harvard-Smithsonian Center for Astrophysics, 60 Garden St., Cambridge MA 02138, USA}
\altaffiltext{\fellow}{Einstein and Spitzer Fellow}

\def\MIT{1}
\def\Princeton{2}
\def\Waterloo{3}
\def\CfA{4}

\def\fellow{$\dagger$}

\begin{document}


\title{Revisiting the Cooling Flow Problem in Galaxies, Groups, and Clusters of Galaxies}

\author{
M.~McDonald\altaffilmark{\MIT},
M.~Gaspari\altaffilmark{\Princeton,\fellow},
B.~R.~McNamara\altaffilmark{\Waterloo}, and
G.~R.~Tremblay\altaffilmark{\CfA}
}

\email{Email: mcdonald@space.mit.edu}   


\begin{abstract}


We present a study of 107 galaxies, groups, and clusters spanning $\sim$3 orders of magnitude in mass, $\sim$5 orders of magnitude in central galaxy star formation rate (SFR), $\sim$4 orders of magnitude in the classical cooling rate (\.{M}$_{cool} \equiv M_{gas}(r<r_{cool})/t_{cool}$) of the intracluster medium (ICM), and $\sim$5 orders of magnitude in the central black hole accretion rate. For each system in this sample we measure the ICM cooling rate, \.{M}$_{cool}$, using archival \emph{Chandra} X-ray data and acquire the SFR and systematic uncertainty in the SFR by combining over 330 estimates from dozens of literature sources. 
With these data, we estimate the efficiency with which the ICM cools and forms stars, finding $\epsilon_{cool} \equiv SFR/\dot{M}_{cool} = 1.4 \pm 0.4$\% for systems with \.M$_{cool}  > 30$ M$_{\odot}$ yr$^{-1}$. For these systems, we measure a slope in the SFR--\.M$_{cool}$ relation greater than unity, suggesting that the systems with the strongest cool cores are also cooling more efficiently. 
We propose that this may be related to, on average, higher black hole accretion rates in the strongest cool cores, which could influence the total amount (saturating near the Eddington rate) and dominant mode (mechanical vs radiative) of feedback.
For systems with \.M$_{cool}  < 30$ M$_{\odot}$ yr$^{-1}$, we find that the SFR and \.{M}$_{cool}$ are uncorrelated, and show that this is consistent with star formation being fueled at a low (but dominant) level by recycled ISM gas in these systems.
We find an intrinsic log-normal scatter in SFR at fixed \.{M}$_{cool}$ of $0.52 \pm 0.06$ dex (1$\sigma$ RMS), suggesting that cooling is tightly self-regulated over very long timescales, but can vary dramatically on short timescales. There is weak evidence that this scatter may be related to the feedback mechanism, with the scatter being minimized ($\sim$0.4 dex) for systems for which the mechanical feedback power is within a factor of two of the cooling luminosity. 

\end{abstract}

%

\section{Introduction}
\setcounter{footnote}{0}

In roughly a third of all galaxy clusters, the central density of the intracluster medium (ICM) is high enough and the central temperature low enough that it ought to cool in a few billion years. This rapidly (in a cosmological sense) cooling region, referred to as a ``cool core'', occupies the inner $\sim$100\,kpc, or $\sim$10\% of the virial radius \citep[e.g.,][]{white97,hudson10,mcdonald17a} and is centered on the brightest cluster galaxy (BCG). Integrating the total ICM mass within this inner region and dividing by the cooling time yields estimates of $\sim$100--1000 M$_{\odot}$ yr$^{-1}$ for the ICM cooling rate for a typical massive galaxy cluster \citep[e.g.][]{white97,peres98,allen01b,hudson10}. Calculations such as this, made shortly after the discovery of the ICM, implied massive ``cooling flows'' of gas falling onto the central BCG in nearly all relaxed clusters \citep[see review by][]{fabian94}. Searches for this gas at cooler temperatures consistently found far less cold gas and young stars than predicted \citep[e.g.,][]{johnstone87,heckman89,mcnamara89,crawford99,donahue00,edge01,edge02,hatch05,odea08,mcdonald10,mcdonald11b,hoffer12,molendi16}, which became known as the ``cooling flow problem''. Stated simply, and summarizing the afore-cited results, the central galaxies in relaxed, cool core clusters appear to be forming new stars at $\sim$1\% of the rate predicted by estimates of the ICM cooling rate.

With the advent of high resolution X-ray imaging from the \emph{Chandra X-ray Observatory} it soon became clear that cool cores were not nearly as relaxed as they had first appeared. These new observations revealed that the ICM in the most relaxed looking clusters is highly dynamic, primarily due to the effects of powerful jets from radio-loud active galactic nuclei. These radio AGN are found at the center of every cool core cluster \citep{sun09b}, and their effect on the ICM can be directly observed via large bubbles in the hot gas, which appear to be inflated by the radio jets \citep[e.g.,][]{birzan04,dunn05,mcnamara05,rafferty06,forman07,mcnamara07,birzan08,birzan12,hlavacek12,hlavacek15}.  These bubbles rise buoyantly to large radius, often allowing an estimate of the duty cycle of AGN feedback when multiple generations of bubbles are observed \citep[e.g.,][]{birzan12}. 
Together with buoyant bubbles, AGN heating is also distributed in the ICM via cocoon shocks and turbulent mixing \citep[][for a brief review]{gaspari13}.
The amount of mechanical energy in these jets is sufficient to offset radiative losses due to cooling, leading to the idea that ``mechanical feedback'' may be responsible for preventing runaway cooling of the ICM \citep[see reviews by][]{mcnamara07,fabian12,mcnamara12}.

At around the same time that the effects of AGN feedback on the ICM were becoming clear, advances in high spectral resolution X-ray and ultraviolet observations revealed a dearth of cooling at low temperatures. Results from the Reflection Grating Spectrograph (RGS) on \emph{XMM-Newton} revealed that the bulk of the ICM cooling was being quenched at temperatures $\sim$1/3 of the ambient core temperature, or roughly $\sim$1 keV for most clusters \citep[see review by][]{peterson06}. These spectroscopic observations set upper limits on the amount of cooling below $\sim$10$^6$\,K at roughly an order of magnitude lower than the classical prediction \citep[e.g.,][]{peterson03,voigt04,peterson06,sanders10,pinto14}. These high quality X-ray observations corroborated early findings from the \emph{FUSE} satellite and more recent findings with the Cosmic Origins Spectrograph on \emph{HST}, which found that the cooling rates through $\sim$10$^{5.5}$\,K (probed via the O\,\textsc{vi} emission line in the far-UV) were ``closer to 30 M$_{\odot}$ yr$^{-1}$ than to the originally suggested values of 10$^2$--10$^3$ M$_{\odot}$ yr$^{-1}$'' \citep[e.g.,][]{bregman05,bregman06,mcdonald14b,donahue17}. Improvements in data quality and analysis, both based on X-ray spectroscopy and far-UV spectroscopy, have supported an emerging picture: the bulk of the ICM cooling is suppressed at high temperatures, but, on average, roughly 10\% of the classical cooling prediction is observed at lower temperatures \citep{mcdonald14b}.

More recently, efforts have shifted away from the question of \emph{how much} cooling is occurring, and have instead focused on what physical conditions lead to, or trigger, the development of cooling instabilities. While the specific details vary, most studies agree that thermally unstable cooling in the ICM develops when the cooling time becomes comparable to or shorter than some characteristic dynamical timescale \citep[e.g.,][]{gaspari12a,mccourt12,sharma12}. Recent numerical works by \cite{gaspari17a} suggest that the turbulent eddy time may represent the timescale of the nonlinear condensation process, while studies by \cite{voit15b} and \cite{li15} advocate for the free-fall time as a dynamical timescale.
\cite{mcnamara16} have suggested thermally unstable cooling ensues when warm gas is lifted outward by rising radio bubbles.  This process would be governed by the infall timescale of a cooling gas parcel, which is bracketed by the free fall time and the timescale set by the terminal speed.
These works all paint a picture of thermally unstable cooling into warm and cold clouds that feed mechanical AGN feedback \citep[e.g., chaotic cold accretion;][]{gaspari17b}.

In this work, we attempt to address the question of how tightly regulated is the cooling--feedback loop in the cores of galaxy clusters. We have assembled a sample of $>$100 galaxy clusters from the literature, spanning $\sim$3 orders of magnitude in mass, $\sim$6 orders of magnitude in black hole accretion rate, $\sim$5 orders of magnitude in cooling rate, and $\sim$5 orders of magnitude in BCG star formation rate (SFR). By approximating the ``cooling efficiency'' as the ratio of the BCG SFR to the ICM cooling rate, we can determine how well AGN are able to prevent runaway cooling in a large ensemble of clusters with a wide variety of properties. In \S2 we will describe the sample selection, which draws from several differently-defined samples in the literature in an attempt to sample a large swath of multi-dimensional parameter space, and how we measure both the SFR and the ICM cooling rate. In \S3 we provide an updated, qualitative assessment of the ``cooling flow problem'', measuring the cooling efficiency for a sample of 107 well-studied systems using the latest data from a wide variety of telescopes.
In \S4, we present a more quantitative examination of the SFR--\.M$_{cool}$ relation, quantifying the slope and scatter as a function of cooling rate. In \S5, we interpret these results for low-mass systems, while in \S6 we consider the opposite end of the mass spectrum, including a discussion of the effects of the black hole accretion rate on the ICM cooling rate. In \S7 we examine the redshift dependence on the cooling flow problem, before summarizing our results and making concluding remarks in \S8.

Throughout this work we assume $\Lambda$CDM cosmology with H$_0$ = 70 km s$^{-1}$ Mpc$^{-1}$, $\Omega_M$ = 0.3, $\Omega_{\Lambda}$ = 0.7. Unless otherwise stated, quoted scatters and uncertainties are 1$\sigma$ RMS. 
\section{Data \& Analysis}

\subsection{Galaxy, Group, and Cluster Samples}

The goal of this work is to compare the maximum cooling rate of the ICM (\.M$_{cool}$) to the observed star formation rate (SFR) in the central galaxy for a large and diverse sample of galaxies, groups, and clusters. The ratio of these two quantities provides an estimate of the cooling efficiency ($\epsilon_{cool} \equiv SFR/\dot{M}_{cool}$) of the hot gas, which is some combination of a hot-phase cooling efficiency (10$^7$K $\rightarrow$ 10$^4$K) and the star formation efficiency. We would like to measure $\epsilon_{cool}$ for a variety of systems, spanning a large range in redshift, mass, cooling rate, and AGN activity, and determine how $\epsilon_{cool}$ scales which each of these quantities.

There is no single sample that spans a suitable range in mass and redshift, while also having measurements of the BCG star formation rate, the AGN activity (e.g., jet power), and available \emph{Chandra} data to measure the cooling rate of the ICM. Instead, we will draw from multiple samples which, when combined or considered individually, will allow us to assess the importance of various properties on the measured value of the cooling efficiency, $\epsilon_{cool}$. These samples, which each contribute an important subset to the total population, are described below in detail.

\subsubsection{Russell \etal (2013): AGN Activity}
One of the primary goals of this work is to study how the suppression of star formation in BCGs depends on the properties of the central AGN. To establish this, we begin with the sample of 57 systems from \cite{russell13}, all of which have estimates of the radiative luminosity of the AGN, the jet power, the black hole mass, and the black hole accretion rate.  These systems span a range in mass from isolated massive elliptical galaxies to rich clusters and, more importantly, include central galaxies with black hole accretion rates ranging from $\sim$10$^{-6}$ to $\sim$1.0 times the Eddington rate. This will allow us to investigate whether the accretion rate of the central AGN (\.M$_{BH}$/\.M$_{edd}$) is linked to the efficiency with which AGN can suppress cooling ($\epsilon_{cool}$).

All sytems in this sample have suitable \emph{Chandra} data to measure cooling rates -- these data were used by \cite{russell13} to measure jet powers (via X-ray cavities) and AGN luminosities. Of these systems, 53 are classified as cool core ($t_{cool,0} < 3$ Gyr; see \S2.2) and 51 of these have sufficiently reliable SFR estimates in the literature (see \S2.3) and will be included in our analysis.

\subsubsection{Cavagnolo \etal (2009): Improved Statistics}
There are a significant number of galaxies, groups, and clusters for which there exists \emph{Chandra} data, and for which we could measure a cooling rate, but were not included in the analysis of \cite{russell13}. In an effort to improve the sample size and, thus, the statistics of any measurements we make, we include all cool core ($t_{cool,0} < 3$ Gyr) groups and clusters from the ACCEPT database \citep{cavagnolo09} that are at $z<0.4$ and that are not already included in the \cite{russell13} sample. This yields an additional 44 systems, all of which have sufficiently deep \emph{Chandra} data to measure a cooling rate. Of these, we were able to obtain reliable SFRs for 33 systems, which we add to our sample. These systems span a large range of mass, providing improved statistics specifically at $\dot{M}_{cool}>100$ M$_{\odot}$ yr$^{-1}$.

\subsubsection{Fogarty \etal (2017): Rare, Massive Systems}

There are relatively few systems in either the \cite{russell13} or \cite{cavagnolo09} samples with $\dot{M}_{cool} \sim 1000$ M$_{\odot}$ yr$^{-1}$. Such systems are rare, corresponding to relaxed clusters with $M_{500} \sim 10^{15}$ M$_{\odot}$ \citep[e.g., the Phoenix cluster;][]{mcdonald12c}. In an effort to populate this extreme end of parameter space in a relatively unbiased way, we include 11 massive clusters from \cite{fogarty17}, which are drawn from the CLASH\footnote{\url{https://archive.stsci.edu/prepds/clash/}} survey. These systems span $\sim$5--30 $\times$ 10$^{14}$ M$_{\odot}$ in mass, have robust SFR estimates based on 16-filter optical-infrared data from \emph{Hubble} \citep{fogarty17}, and have deep \emph{Chandra} data  \citep{donahue14}. These high masses correspond to high cooling rates, spanning $\dot{M}_{cool}\sim300-2000$ M$_{\odot}$ yr$^{-1}$ for this 11-cluster sample. The inclusion of these rare, massive systems improves our understanding of the scatter in SFR at fixed cooling rate for the most massive (and most rapidly cooling) systems.

In addition to these 11 systems, we further include the Perseus \citep[e.g.,][]{fabian03,canning14} and Phoenix \citep{mcdonald12c} clusters, which have extreme cooling and star formation rates but are not included in the \cite{russell13} or ACCEPT samples.

\subsubsection{Fraser-McKelvie \etal (2014): Completeness}

One potentially large source of bias in this analysis is that we rely on the literature to provide estimates of the BCG SFR. This could lead to a bias towards ``exciting'' systems (those with high $\epsilon_{cool}$), which are likely overrepresented in the literature. Further, we suspect that many non-detections are missing from the literature, which means that a literature search for the SFR of a given system will inevitably be biased high, especially as we approach typical sensitivity limits ($\lesssim$0.1 M$_{\odot}$ yr$^{-1}$).

To determine how important these biases are, we include a luminosity-complete sample from \cite{fraser14}. This analysis considered a volume-limited ($z<0.1$) sample of galaxy groups and clusters that was complete above L$_{X} > 10^{44}$ erg s$^{-1}$. Unfortunately, not all of these systems have suitable \emph{Chandra} data with which we can measure the cooling rate. To maximize completeness, we make a luminosity cut at L$_{X} > 3.3\times 10^{44}$ erg s$^{-1}$, above which the sample is maximally represented in the \emph{Chandra} archive, with $>$93\% having suitable X-ray data. This sample has a total of 33 cool cores, 12 of which are not yet included in this sample. When considering selection biases, we will isolate this sample of 33 groups and clusters, which is representative of the true cluster population within $z<0.1$.

In total, the sample comprises 107 galaxies, groups, and clusters drawn from several literature sources. Within this large, inhomogeneous sample are several subsamples which allow us to examine trends in $\epsilon_{cool}$ as a function of AGN and cluster properties, and to assess the systematic biases in our literature-based selection.

\subsection{Cooling Rates}

We define the ``classical'' cooling rate in this work in a straightforward and easy-to-calculate way,

\begin{equation}
\dot{M}_{cool} = \frac{M_{gas}(r<r_{cool})}{t_{cool}} ,
\end{equation}

\noindent{}where $r_{cool}$ is the radius within which the cooling time is less than 3\,Gyr, and $t_{cool}$ is defined to be 3\,Gyr. While this quantity is often derived with respect to a cooling time of 7.7\,Gyr (the time since $z=1$), we choose instead a shorter timescale to more closely probe the active cooling. In \cite{mcdonald10,mcdonald11a}, we showed that thermal instabilities (traced by H$\alpha$-emitting filaments) extend to a radius within which the cooling time is $\lesssim$3\,Gyr in the most extended cases, such as Perseus, Abell~1795, Abell~2597, Sersic~159-03, and NGC~4325. Thus, this choice of radius allows us to probe the maximum cooling rate over a volume where there is evidence that cooling actually occurs.

To calculate the cooling rate, we require density and temperature profiles over a large radial range. These are available for many of our systems from the ACCEPT\footnote{https://web.pa.msu.edu/astro/MC2/accept/} database \citep{cavagnolo09}. We note that, while it has recently been found that ACCEPT profiles may be biased in the innermost bins \citep{lakhchaura16}, this does not have a large effect on \.{M}$_{cool}$ which is measured, on average, on scales of $\sim$50--100 kpc. Further, we have re-fit the thermodynamic profiles provided by the ACCEPT collaboration with the analytic profiles described in \cite{vikhlinin06a} for both the temperature and density profiles. We have compared these recomputed \.{M}$_{cool}$ values to those from \cite{vikhlinin09a} for systems overlapping between the two samples, finding an average difference of $10.0\pm9.6$\%, suggesting that the ACCEPT data products are of sufficient quality for this work. 

A total of 70 systems had density and temperature profiles available from ACCEPT. For the remaining 37 systems, we obtained \emph{Chandra} data from the archive and measured deprojected density and temperature profiles, following the methodology laid out in \cite{vikhlinin06a} and using the latest version of CIAO (v4.7) and CALDB (v4.7.3) at the time of writing. Once deprojected thermodynamic profiles were extracted, these data were treated in the same way as the ACCEPT profiles, allowing for a uniform analysis.

Given a deprojected density and temperature profile, we calculate the cooling rate assuming

\begin{equation}
t_{cool} = \frac{3}{2}\frac{(n_e+n_p)kT}{n_en_H\Lambda(T,Z)} ,
\end{equation}

\noindent{}where we assume $n_p=0.92n_e$, $n_H=0.83n_e$, and $\Lambda(T,Z)$ is the cooling function assuming $Z_{\odot}/3$ metallicity \citep{tozzi01}. The cooling time profile is calculated in this way for each of the 107 galaxies, groups, and clusters in our sample, from which we can estimate the radius at which $t_{cool} = 3$ Gyr. Integrating the gas density within a sphere bounded by this radius provides an estimate of the total mass available for cooling, from which we can derive the maximum cooling rate from Equation 1. We calculate 1,000 cooling time and cooling rate profiles assuming Gaussian errors on the temperature and density profiles, allowing for a Monte Carlo calculation of the uncertainties on the cooling time, $t_{cool}$, the cooling radius, $r_{cool}$, and the classical cooling rate, \.M$_{cool}$.

We note that this cooling rate approximates the maximum allowed cooling rate for each systems, and should not be confused with the \emph{spectroscopic} cooling rate \citep[see e.g.,][]{peterson03,peterson06,molendi16}, which measures the amount of gas that is, in fact, cooling through some specific temperature or gas phase.

\subsection{Star Formation Rates}

Our goal in assembling SFRs for each of the BCGs in the sample is to provide a reliable estimate based on either a variety of methods or a single, robust method, to assess the systematic uncertainties in the estimate, and to avoid, wherever possible, contamination from AGN. To this end, we first isolate systems for which we either know or suspect the presence of an AGN, which could contaminate the SFR estimate. This characterization is made based on a combination of a literature search for each system as well as a measurement of the W1-W2 color from the WISE mission \citep{stern05, wright10}. Based on the mid-IR color criteria of \cite{stern05}, we find evidence for an AGN which may contaminate the estimate of SFR in the following systems: H1821+643, IRAS~09104+4109, Cygnus~A, RBS~797, 3C295, 3C388, Zw2089, Abell~1068, and Abell~2667. For each of these systems, we require an estimate of the SFR that either spatially or spectrally separates the AGN and starburst component. We were unable to find such well-measured SFRs for 3C295 or 3C388, so we will refer to SFRs for these systems as upper limits. For the remaining 7 systems, along with Phoenix and Perseus, we obtained either spatially decomposed \citep[Abell~611, RBS797;][]{cavagnolo11,donahue15} or spectrally decomposed \citep[H1821+643, 3C186, IRAS~09104+4109, Abell~2667, Cygnus~A, Zw2089, Phoenix, Perseus;][]{privon12,rawle12,ruiz13,mittal15,podigachoski15,mittal17} SFR estimates. 

\begin{deluxetable}{ccccl}[htb]
\tabletypesize{\footnotesize} 
\tablecolumns{4}
\tablewidth{0pt}
\tablecaption{Data for Galaxies, Groups, and Clusters \\from Russell \etal (2013)}
\tablehead{
\colhead{Name} & \colhead{$\log_{10}$(\.M$_{cool}$)} & \colhead{$\log_{10}$(SFR)} & \colhead{Ref.}\\
 & \colhead{[M$_{\odot}$/yr]} & \colhead{[M$_{\odot}$/yr]} & }
 \startdata
2A0335 & $+$2.26 $\pm$ 0.01 & $-$0.33 $\pm$  0.18 & $^{*cgh}$ \\
ABELL 0085 & $+$1.94 $\pm$ 0.01 & $-$1.03 $\pm$  1.47 & $^{*bcdehlo}$ \\
ABELL 0133 & $+$1.79 $\pm$ 0.01 & $-$0.63 $\pm$  0.88 & $^{*ceh}$ \\
ABELL 0262 & $+$0.48 $\pm$ 0.04 & $-$0.65 $\pm$  0.06 & $^{*bhl}$ \\
ABELL 0478 & $+$2.64 $\pm$ 0.01 & $+$0.29 $\pm$  0.04 & $^{ceghl}$ \\
ABELL 1795 & $+$2.27 $\pm$ 0.02 & $+$0.54 $\pm$  0.70 & $^{*ceghlnq}$ \\
ABELL 1835 & $+$3.07 $\pm$ 0.06 & $+$2.07 $\pm$  0.20 & $^{*cghjoqs}$ \\
ABELL 2029 & $+$2.43 $\pm$ 0.05 & $-$0.06 $\pm$  0.05 & $^{*dln}$ \\
ABELL 2052 & $+$1.66 $\pm$ 0.02 & $-$0.36 $\pm$  0.53 & $^{*bcdehlns}$ \\
ABELL 2199 & $+$1.68 $\pm$ 0.05 & $+$0.09 $\pm$  0.93 & $^{*chlnqs}$ \\
ABELL 2390 & $+$2.18 $\pm$ 0.06 & $+$1.04 $\pm$  0.33 & $^{*chjs}$ \\
ABELL 2597 & $+$2.49 $\pm$ 0.05 & $+$0.60 $\pm$  0.36 & $^{*cdghjlq}$ \\
ABELL 4059 & $+$1.09 $\pm$ 0.06 & $-$0.55 $\pm$  0.67 & $^{bchl}$ \\
CENTAURUS & $+$0.97 $\pm$ 0.01 & $-$0.79 $\pm$  0.12 & $^{*ch}$ \\
HCG 62 & $+$0.73 $\pm$ 0.02 & $-$1.20 $\pm$  1.28 & $^{*ch}$ \\
HERCULES A & $+$1.75 $\pm$ 0.01 & $-$0.44 $\pm$  0.53 & $^{*cdh}$ \\
HYDRA A & $+$2.04 $\pm$ 0.02 & $+$0.61 $\pm$  0.45 & $^{*cdeghq}$ \\
M84 & $-$0.89 $\pm$ 0.03 & $-$1.24 $\pm$  0.40 & $^{m}$ \\
M87 & $+$1.29 $\pm$ 0.00 & $-$0.85 $\pm$  1.14 & $^{*chn}$ \\
MKW3S & $+$1.36 $\pm$ 0.05 & $-$0.55 $\pm$  0.42 & $^{*cdlnos}$ \\
MS0735 & $+$2.42 $\pm$ 0.08 & $+$0.52 $\pm$  0.25 & $^{*cg}$ \\
NGC0507 & $+$0.78 $\pm$ 0.05 & $-$0.64 $\pm$  0.12 & $^{*hm}$ \\
NGC1316 & $-$0.51 $\pm$ 0.01 & $-$0.05 $\pm$  0.33 & $^{*mn}$ \\
NGC1600 & $-$0.30 $\pm$ 0.04 & $-$0.74 $\pm$  0.39 & $^{*a}$ \\
NGC4261 & $-$0.51 $\pm$ 0.01 & $-$0.70 $\pm$  0.32 & $^{*m}$ \\
NGC4472 & $-$0.01 $\pm$ 0.00 & $-$1.20 $\pm$  0.34 & $^{*am}$ \\
NGC4636 & $-$0.42 $\pm$ 0.06 & $-$1.62 $\pm$  0.27 & $^{*ahm}$ \\
NGC4782 & $+$0.23 $\pm$ 0.03 & $-$0.67 $\pm$  0.37 & $^{*a}$ \\
NGC5044 & $+$1.94 $\pm$ 0.03 & $-$0.66 $\pm$  0.12 & $^{*ahm}$ \\
NGC5813 & $+$0.34 $\pm$ 0.00 & $-$1.35 $\pm$  0.13 & $^{*am}$ \\
NGC5846 & $+$0.22 $\pm$ 0.01 & $-$1.06 $\pm$  0.22 & $^{*ahmo}$ \\
NGC6269 & $+$0.09 $\pm$ 0.11 & $-$0.38 $\pm$  0.40 & $^{*}$ \\
NGC6338 & $+$0.88 $\pm$ 0.01 & $-$0.37 $\pm$  0.62 & $^{*nos}$ \\
PKS 0745-191 & $+$2.89 $\pm$ 0.01 & $+$1.13 $\pm$  0.24 & $^{*bcghq}$ \\
ABELL 3581 & $+$1.35 $\pm$ 0.22 & $-$0.11 $\pm$  0.48 & $^{*c}$ \\
RXC J0352.9+1941 & $+$2.30 $\pm$ 0.03 & $+$0.75 $\pm$  0.40 & $^{*}$ \\
RXC J1459.4-1811 & $+$2.48 $\pm$ 0.04 & $+$1.59 $\pm$  0.40 & $^{*}$ \\
RXC J1524.2-3154 & $+$2.23 $\pm$ 0.01 & $+$0.61 $\pm$  0.40 & $^{*}$ \\
RXC J1558.3-1410 & $+$2.10 $\pm$ 0.03 & $+$0.74 $\pm$  0.40 & $^{*}$ \\
Sersic 159-03 & $+$2.37 $\pm$ 0.02 & $+$0.01 $\pm$  0.41 & $^{*cehln}$ \\
Zw 2701 & $+$1.81 $\pm$ 0.27 & $-$0.46 $\pm$  0.49 & $^{chos}$ \\
Zw 3146 & $+$2.87 $\pm$ 0.11 & $+$1.84 $\pm$  0.34 & $^{*cdhjoqs}$
\enddata
\tablecomments{See \S2.1.1 for a description of this sample. References are:
$^*$:GALEX+WISE SFRs derived following \cite{hao11};
$^a$:\cite{macchetto96}; 
$^b$:\cite{odea08}; 
$^c$:\cite{cavagnolo09}; 
$^d$:\cite{hicks10}; 
$^e$:\cite{mcdonald10};
$^f$:\cite{cavagnolo11};
$^g$:\cite{donahue11};
$^h$:\cite{hoffer12};
$^i$:\cite{privon12};
$^j$:\cite{rawle12};
$^k$:\cite{ruiz13};
$^l$:\cite{fraser14};
$^m$:\cite{amblard14}
$^n$:\cite{bai15};
$^o$:\cite{chang15};
$^p$:\cite{donahue15};
$^q$:\cite{mittal15};
$^r$:\cite{podigachoski15};
$^s$:\cite{salim16};
$^t$:\cite{mittal17}
}
\label{table:data1}
\end{deluxetable}

For the 11 massive clusters in the sample of \cite{fogarty17}, we use the published SFRs. These  SFRs are derived based on 16+ band \emph{Hubble Space Telescope} imaging, with careful modeling of the underlying stellar populations -- these estimates are significantly more secure than typical BCG SFRs and include a careful estimate of the systematic uncertainties in stellar populations and internal extinction. 

For systems in the \cite{russell13} and ACCEPT samples with no obvious AGN contamination, which comprise the bulk of this sample (75/107), we determine SFRs based on an ensemble of measurements from the literature. For the vast majority of these measurements, the reported uncertainty reflects only the measurement uncertainty on the single flux used to derive the SFR (e.g., $f_{H\alpha}$, $f_{UV}$, $f_{24\mu m}$, etc), and does not include uncertainty in, for example, ionization sources, dust extinction, dust emission, AGN contamination, etc. In an effort to properly assess the uncertainty on the SFR in these systems, we acquire multiple measurements for each system from the literature, based on different techniques/data, and measure the scatter in these measurements. Specifically, for each system we measure the logarithmic mean and the log-normal scatter in the SFR from the available literature measurements. This measured scatter, which is used as the ``real'' uncertainty on the SFR, is more representative of the difficulty in constraining the SFR, given uncertainty in the initial mass function, intrinsic extinction, etc.
We obtained 302 SFR estimates for these 75 systems from the literature, based on IR dust emission \citep{odea08,hoffer12,rawle12,fraser14}, UV stellar continuum \citep{hicks10,hoffer12,bai15}, emission lines \citep{macchetto96,cavagnolo09,mcdonald10,donahue11}, and optical SED fitting \citep{amblard14,chang15,mittal15,salim16}. In addition, we infer the dust-corrected SFR based on the combination of archival GALEX UV continuum and mid-IR WISE emission for 72 of these systems, following \cite{hao11}, which provides an additional non-literature estimate for the bulk of our systems. This leads to a total input of 374 SFR measurements for 75 systems, or an average of 5 independent measurements per system, allowing us to constrain the measurement scatter.

Finally, for clusters in the \cite{fraser14} sample, we recompute SFRs based on the 12$\mu$m flux. As discussed in \cite{green16}, the SFRs published in this work lack an important k-correction, which leads to systematically biased measurements. Beginning with the published photometry in \cite{fraser14} and following the methodology in \cite{green16}, we compute the expected 12$\mu$m flux for a passive population as a function of redshift, and subtract this from the observed 12$\mu$m emission for each BCG. These continuum-subtracted 12$\mu$m fluxes are then converted to SFRs following \cite{cluver14}. This re-analysis leads to a significantly higher number of non-detections than were presented in \cite{fraser14}. These SFRs, including measured upper limits, will be used for an assessment of biases in the literature.

In Tables \ref{table:data1}--\ref{table:data5} we provide the classical cooling rate and BCG SFR for each cluster in our sample for which we have made this measurement or obtained it from the literature.

\begin{deluxetable}{ccccl}
\tabletypesize{\footnotesize} 
\tablecolumns{4}
\tablewidth{0pt}
\tablecaption{Data for Groups and Clusters from ACCEPT Sample}
\tablehead{
\colhead{Name} & \colhead{$\log_{10}$(\.M$_{cool}$)} & \colhead{$\log_{10}$(SFR)} & \colhead{Ref.}\\
 & \colhead{[M$_{\odot}$/yr]} & \colhead{[M$_{\odot}$/yr]} & }
 \startdata
ABELL 0496 & $+$1.75 $\pm$ 0.03 & $-$0.71 $\pm$  0.03 & $^{*cel}$ \\
ABELL 0963 & $+$1.51 $\pm$ 3.44 & $+$0.12 $\pm$  0.25 & $^{*hs}$ \\
ABELL 1204 & $+$2.60 $\pm$ 0.03 & $+$0.17 $\pm$  0.48 & $^{*bcdhs}$ \\
ABELL 1361 & $+$1.66 $\pm$ 0.19 & $+$0.30 $\pm$  0.45 & $^{chs}$ \\
ABELL 1413 & $+$1.74 $\pm$ 0.12 & $+$0.28 $\pm$  0.40 & $^{*h}$ \\
ABELL 1644 & $+$0.69 $\pm$ 0.07 & $-$0.36 $\pm$  0.32 & $^{*cehl}$ \\
ABELL 1650 & $+$1.46 $\pm$ 0.08 & $-$1.61 $\pm$  0.12 & $^{el}$ \\
ABELL 1664 & $+$2.21 $\pm$ 0.04 & $+$1.12 $\pm$  0.04 & $^{*bh}$ \\
ABELL 1689 & $+$2.31 $\pm$ 0.10 & $+$1.04 $\pm$  0.40 & $^{h}$ \\
ABELL 1991 & $+$1.58 $\pm$ 0.04 & $-$0.15 $\pm$  0.72 & $^{*cehls}$ \\
ABELL 2107 & $+$0.02 $\pm$ 0.31 & $-$0.37 $\pm$  0.07 & $^{*hls}$ \\
ABELL 2142 & $+$1.73 $\pm$ 0.07 & $-$0.64 $\pm$  0.98 & $^{*dehls}$ \\
ABELL 2151 & $+$0.80 $\pm$ 0.02 & $-$0.25 $\pm$  0.47 & $^{*hs}$ \\
ABELL 2204 & $+$2.70 $\pm$ 0.01 & $+$0.90 $\pm$  0.16 & $^{*bh}$ \\
ABELL 2244 & $+$1.47 $\pm$ 0.02 & $-$0.66 $\pm$  0.40 & $^{l}$ \\
ABELL 2261 & $+$2.10 $\pm$ 0.10 & $+$0.76 $\pm$  0.34 & $^{hn}$ \\
ABELL 2556 & $+$2.15 $\pm$ 0.08 & $-$0.27 $\pm$  0.40 & $^{*h}$ \\
ABELL 2626 & $+$1.21 $\pm$ 0.06 & $-$0.63 $\pm$  0.61 & $^{*chl}$ \\
ABELL 3112 & $+$1.93 $\pm$ 0.05 & $-$0.04 $\pm$  0.49 & $^{*bcdhl}$ \\
ABELL 3528S & $+$0.97 $\pm$ 0.03 & $+$0.13 $\pm$  0.72 & $^{*hln}$ \\
ABELL 3581 & $+$1.35 $\pm$ 0.22 & $-$0.18 $\pm$  0.57 & $^{*ch}$ \\
AWM7 & $+$0.58 $\pm$ 0.01 & $-$0.53 $\pm$  0.30 & $^{*hl}$ \\
RX J0439+0520 & $+$2.39 $\pm$ 0.23 & $+$0.95 $\pm$  0.30 & $^{bch}$ \\
RX J1000.4+4409 & $+$0.91 $\pm$ 0.42 & $-$0.82 $\pm$  0.23 & $^{cos}$ \\
RX J1320.2+3308 & $-$0.41 $\pm$ 5.59 & $-$0.85 $\pm$  1.31 & $^{cs}$ \\
RX J1504.1-0248 & $+$3.29 $\pm$ 0.08 & $+$1.93 $\pm$  0.10 & $^{*hqs}$ \\
RX J1539.5-8335 & $+$2.19 $\pm$ 0.05 & $+$0.27 $\pm$  0.06 & $^{*l}$ \\
RX J1720.1+2638 & $+$2.63 $\pm$ 0.03 & $+$0.41 $\pm$  0.59 & $^{*chjs}$ \\
RX J2129.6+0005 & $+$2.36 $\pm$ 0.33 & $+$0.51 $\pm$  0.29 & $^{*bhjns}$ \\
MS 1455.0+2232 & $+$2.78 $\pm$ 0.02 & $+$1.03 $\pm$  0.31 & $^{*dhos}$ 
\enddata
\tablecomments{See \S2.1.2 for a description of this sample. Cool core clusters in ACCEPT that are included in the sample of \cite{russell13} have been excluded here. References are:
$^*$:GALEX+WISE SFRs derived following \cite{hao11};
$^a$:\cite{macchetto96}; 
$^b$:\cite{odea08}; 
$^c$:\cite{cavagnolo09}; 
$^d$:\cite{hicks10}; 
$^e$:\cite{mcdonald10};
$^f$:\cite{cavagnolo11};
$^g$:\cite{donahue11};
$^h$:\cite{hoffer12};
$^i$:\cite{privon12};
$^j$:\cite{rawle12};
$^k$:\cite{ruiz13};
$^l$:\cite{fraser14};
$^m$:\cite{amblard14}
$^n$:\cite{bai15};
$^o$:\cite{chang15};
$^p$:\cite{donahue15};
$^q$:\cite{mittal15};
$^r$:\cite{podigachoski15};
$^s$:\cite{salim16};
$^t$:\cite{mittal17}
}
\label{table:data2}
\end{deluxetable}

\begin{deluxetable}{ccccl}
\tabletypesize{\footnotesize} 
\tablecolumns{4}
\tablewidth{0pt}
\tablecaption{Data for Groups and Clusters with Bright Central AGN}
\tablehead{
\colhead{Name} & \colhead{$\log_{10}$(\.M$_{cool}$)} & \colhead{$\log_{10}$(SFR)} & \colhead{Ref.}\\
 & \colhead{[M$_{\odot}$/yr]} & \colhead{[M$_{\odot}$/yr]} & }
 \startdata
3C295 & $+$2.98 $\pm$ 0.04 &  $<$$+$1.53 & $^a$ \\ 
3C388 & $+$0.18 $\pm$ 0.31 &  $<$$-$0.03 & $^a$ \\ 
ABELL 1068 & $+$2.55 $\pm$ 0.02 & $+$1.06 $\pm$  0.36 & $^f$\\ 
ABELL 2667 & $+$2.76 $\pm$ 0.12 & $+$0.94 $\pm$  0.01 &  $^c$\\ 
H1821+643 & $+$2.65 $\pm$ 0.05 & $+$2.65 $\pm$  0.33 & $^e$ \\ 
IRAS 09104+4109 & $+$3.01 $\pm$ 0.04 & $+$2.49 $\pm$  0.39 & $^e$\\ 
CYGNUS A & $+$2.15 $\pm$ 0.01 & $+$1.60 $\pm$  0.33 & $^d$\\ 
Perseus & $+$2.67 $\pm$ 0.05 & $+$1.85 $\pm$ 0.28 & $^f$\\ 
Phoenix &  $+$3.23 $\pm$ 0.08 & $+$2.79 $\pm$ 0.36 & $^g$\\ 
RBS797 & $+$3.15 $\pm$ 0.24 & $+$0.78 $\pm$  0.29 & $^b$\\ 
Zw 2089 & $+$2.61 $\pm$ 0.04 & $+$1.31 $\pm$  0.02 & $^c$ 
\enddata
\tablecomments{See \S2.1.1 and \S2.3 for a description of this sample. References are:
$^a$:\cite{shi07}; $^b$:\cite{cavagnolo11}; $^c$:\cite{rawle12}; $^d$:\cite{privon12}; $^e$:\cite{ruiz13}; $^f$:\cite{mittal15}; $^g$:\cite{mittal17}}
\label{table:data3}
\end{deluxetable}

\begin{deluxetable}{ccccl}
\tabletypesize{\footnotesize} 
\tablecolumns{4}
\tablewidth{0pt}
\tablecaption{Data for Massive, Rare Clusters}
\tablehead{
\colhead{Name}& \colhead{$\log_{10}$(\.M$_{cool}$)} & \colhead{$\log_{10}$(SFR)} & \colhead{Ref.}\\
 & \colhead{[M$_{\odot}$/yr]} & \colhead{[M$_{\odot}$/yr]} & }
 \startdata
A383 & $+$2.47 $\pm$ 0.04 & $+$0.18 $\pm$  0.24 & F17 \\
MACS0329.7-0211 & $+$2.75 $\pm$ 0.04 & $+$1.60 $\pm$  0.19 & F17 \\
MACS0429.6-0253 & $+$2.72 $\pm$ 0.04 & $+$1.53 $\pm$  0.23 & F17 \\
MACS1115.8+0129 & $+$2.81 $\pm$ 0.04 & $+$0.85 $\pm$  0.28 & F17 \\
MACS1423.8+2404 & $+$2.85 $\pm$ 0.02 & $+$1.41 $\pm$  0.18 & F17 \\
MACS1720.3+3536 & $+$2.71 $\pm$ 0.05 & $+$0.19 $\pm$  0.26 & F17 \\
MACS1931.8-2634 & $+$3.03 $\pm$ 0.10 & $+$2.42 $\pm$  0.20 & F17 \\
MS 2137.3-2353 & $+$2.78 $\pm$ 0.02 & $+$0.25 $\pm$  0.29 & F17 \\
MACS1347.5-1144 & $+$3.01 $\pm$ 0.08 & $+$1.07 $\pm$  0.23 & F17 \\
RXJ1532.9+3021 & $+$3.03 $\pm$ 0.04 & $+$1.99 $\pm$  0.19 & F17 
\enddata
\tablecomments{See \S2.1.3 for a description of this sample. References are: F17: \cite{fogarty17}}
\label{table:data4}
\end{deluxetable}

\begin{deluxetable}{ccccl}
\tabletypesize{\footnotesize} 
\tablecolumns{4}
\tablewidth{0pt}
\tablecaption{Data for Volume-Complete Sample of Galaxy Groups and Clusters}
\tablehead{
\colhead{Name} & \colhead{$\log_{10}$(\.M$_{cool}$)} & \colhead{$\log_{10}$(SFR)} & \colhead{Ref.}\\
 & \colhead{[M$_{\odot}$/yr]} & \colhead{[M$_{\odot}$/yr]} & }
 \startdata
ABELL 0550 & $+$0.51 $\pm$ 0.27 & $<$$-$1.62 & F14, G16 \\
ABELL 1651 & $-$0.10 $\pm$ 0.10 & $<$$-$0.51 & F14, G16 \\
ABELL 2110 & $+$1.68 $\pm$ 0.07 & $<$$-$0.15 & F14, G16 \\
ABELL 2249 & $+$0.13 $\pm$ 0.28 & $-$1.66 $\pm$  0.40 & F14, G16 \\
ABELL 2426 & $+$1.66 $\pm$ 0.12 & $<$$-$0.41 & F14, G16 \\
ABELL 3571 & $+$0.46 $\pm$ 0.15 & $<$$-$0.47 & F14, G16 \\
ABELL 3911 & $-$0.02 $\pm$ 0.56 & $-$0.88 $\pm$  0.29 & F14, G16 \\
ABELL 3921 & $-$1.00 $\pm$ 1.22 & $-$0.29 $\pm$  0.05 & F14, G16 \\
NRGB045 & $+$0.59 $\pm$ 0.06 & $-$1.29 $\pm$  0.21 & F14, G16\\
RX J2218.0-6511 & $+$1.56 $\pm$ 0.03 & $-$0.82 $\pm$  0.18 & F14, G16 \\
RX J2223.9-0137 & $+$1.30 $\pm$ 0.04 & $<$$-$1.46 & F14, G16 \\
ZWCL1742.1+3306 & $-$0.36 $\pm$ 0.05 & $-$0.11 $\pm$  0.12 & F14, G16
\enddata
\tablecomments{Groups and clusters in the volume-complete sample that appear in Tables \ref{table:data1}--\ref{table:data4} have been excluded here. See \S2.1.4 for a description of the sample. Wise band 3 photometry was acquired from \cite{fraser14}, and star formation rates recomputed following \cite{green16}, as described in \S2.3}.
\label{table:data5}
\end{deluxetable}


\section{The Relationship Between the ICM Cooling Rate and the BCG Star Formation Rate}

\begin{figure}[htb]
\centering
\includegraphics[width=0.49\textwidth]{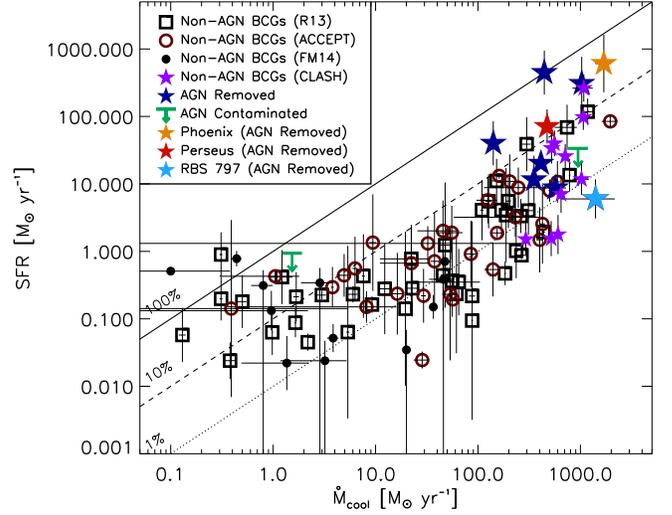}
\caption{Star formation rate of the central galaxy as a function of the predicted ICM cooling rate, for 107 galaxies, groups, and clusters described in \S2. Open squares and open circles show ensemble measurements of the SFR for systems in the \cite{russell13} and \cite{cavagnolo09} samples, respectively, where the uncertainty represents the scatter in measurements from different literature sources. Blue stars show systems from these two surveys that have AGN that contaminate their star formation rate estimates -- for these systems we show single literature values where the AGN and starburst component have been modeled simultaneously. Green arrows show upper limits for systems for which the AGN could not be removed, while purple stars show systems from the CLASH survey \citep{fogarty17}. Closed black circles show SFR estimates from \cite{fraser14}. Diagonal lines show the 1\%, 10\%, and 100\% lines -- the bulk of the systems shown here have star formation rates that are 1--10\% of their predicted cooling rate.}
\label{fig:sfr_combine}
\end{figure}

In Figure \ref{fig:sfr_combine}, we show the measured star formation rates and the classical cooling rates for the full sample of 107 galaxies, groups, and clusters described in \S2. This figure reveals several interesting trends, both old and new. First, we find that there are no systems for which the BCG has a SFR significantly higher than the predicted ICM cooling rate, suggesting that the latter represents an upper limit on the former. The overall distribution of points does not appear to be well-described by a single slope, suggesting multiple physical processes at play.
At the low-\.M$_{cool}$ end, the star formation rate appears to be uncorrelated with the cooling rate -- we will investigate this further in \S4. At cooling rates above $\sim$10 M$_{\odot}$ yr$^{-1}$, the SFR is correlated with the cooling rate, with typical ratios between $\sim$1--10\%. We will refer to this ratio as $\epsilon_{cool}$, or the efficiency with which the hot gas cools and forms stars. For systems at the high-\.M$_{cool}$ end, we see a potential increase in $\epsilon_{cool}$, with the bulk of systems at \.M$_{cool}$ $\sim$ 1000 M$_{\odot}$ yr$^{-1}$ having $\epsilon_{cool}$ $\sim$ 0.1--1.0. We will investigate this trend further in \S5. We find no deviations from the overall trend between samples, with the R13, ACCEPT, and CLASH samples all overlapping.

Over a narrow range in cooling rates around $\dot{M}_{cool}\sim500$ M$_{\odot}$ yr$^{-1}$, \cite{fogarty15} showed that there were two orders of magnitude of scatter in BCG star formation rates for massive clusters in the CLASH sample -- this trend is replicated in Figure \ref{fig:sfr_combine}, with well-studied starburst-BCG systems such as Perseus and Phoenix lying at the extreme end of this distribution. At the opposite end (low-SFR) of this distribution are less well-studied systems, such as RBS797 and MS2137-2353, which have classical cooling rates of $\sim$1000 M$_{\odot}$ yr$^{-1}$ but star formation rates of only $\sim$2 M$_{\odot}$ yr$^{-1}$.  We will investigate further in \S5 the differences between these highly efficient and highly inefficient cooling systems.

\begin{figure}[htb]
\centering
\includegraphics[width=0.49\textwidth]{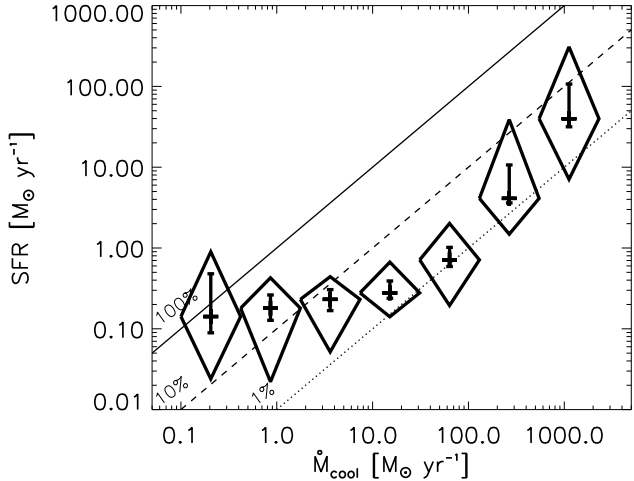}
\caption{The same data as Figure \ref{fig:sfr_combine}, but logarithmically binned in cooling rate. This figure demonstrates the roughly uniform slope for systems with $\dot{M}_{cool} > 10$ M$_{\odot}$ yr$^{-1}$ and the flattening at $\dot{M}_{cool} < 10$M$_{\odot}$ yr$^{-1}$. For all points, the 1$\sigma$ scatter is enclosed by the diamond, while the median value and the uncertainty on the median, is shown by the black cross. These data support a picture in which cooling is suppressed by, on average, a factor of $\sim$50 in massive systems. This suppression factor ranges, on a system-by-system basis, from several hundred (fully suppressed), to as low as 0 (unsuppressed).}
\label{fig:sfr_dmdt_averages}
\end{figure}

In Figure \ref{fig:sfr_dmdt_averages}, we show the data from Figure \ref{fig:sfr_combine}, binned by classical cooling rate in equally-spaced logarithmic bins. This figure makes more clear the strong correlation between star formation rate and cooling rate at $\dot{M}_{cool} > 10$ M$_{\odot}$ yr$^{-1}$, and the flattening of the trend at $\dot{M}_{cool} < 10$ M$_{\odot}$ yr$^{-1}$. 
In the four bins with $\dot{M}_{cool} > 10$ M$_{\odot}$ yr$^{-1}$, which span more than two orders of magnitude in cooling rate, we measure median star formation rates corresponding to 1.8\%, 1.1\%, 1.5\%, and 3.6\% of the cooling rate. This tight correlation implies that the process responsible for regulating cooling is equally capable of quenching small amounts of cooling and massive cooling flows. 
For systems with $\dot{M}_{cool} < 10$ M$_{\odot}$ yr$^{-1}$, we measure only a small range in the median star formation rates (0.14--0.23 M$_{\odot}$ yr$^{-1}$), despite a factor of 20 difference in cooling rates. This implies one of several scenarios: that cooling is becoming more efficient in low-mass halos, that there is a non-ICM origin for the cool gas fueling star formation, that we are missing a population of low-mass systems with star formation rates $<$0.01 M$_{\odot}$ yr$^{-1}$, or that the star formation rates are being overestimated at the low end. We will investigate these possibilities further in \S4.

\begin{figure}[htb]
\centering
\includegraphics[width=0.49\textwidth]{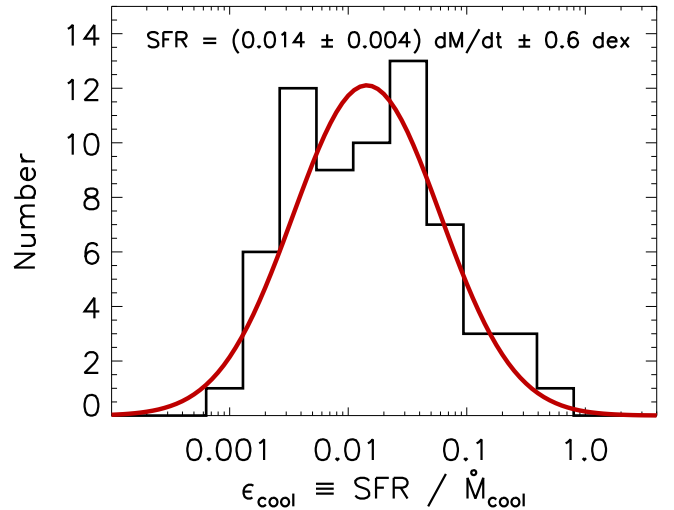}
\caption{Ratio of the star formation rate in the central BCG to the predicted ICM cooling rate for 75 groups and clusters with $\dot{M}_{cool} > 30$ M$_{\odot}$ yr$^{-1}$. The distribution of measured ``cooling efficiencies'' ($\epsilon_{cool}$) is well-modeled by a log-normal distribution that peaks at $1.4\pm0.4$\% and has a width of 0.6 dex.}
\label{fig:effhist}
\end{figure}

Given that the slope in Figure \ref{fig:sfr_dmdt_averages} appears to be roughly uniform for systems with $\dot{M}_{cool} > 30$ M$_{\odot}$ yr$^{-1}$, we can consider the distribution of ``cooling efficiencies'' ($\epsilon_{cool}$, the star formation rate normalized to the cooling rate) for this subsample of systems.  In Figure \ref{fig:effhist}, we show that the distribution of $\epsilon_{cool}$ is log-normal, with a peak at $1.4\pm0.4$\%. This is fully consistent with the canonical ``1\%'' that is typically quoted. This efficiency would, of course, be slightly lower if we considered the more widely used cooling radius defined by a cooling time of 7.7 Gyr. 
The width of this distribution, $\sim$0.6 dex, is quite high -- at a fixed cooling rate, it implies BCG star formation rates spanning more than three orders of magnitude for samples of $>$100 clusters. 
This suggests that the coupling between AGN feedback and cooling is far from perfect, and that BCGs in cool cores likely experience periods of highly efficient star formation followed by periods of quenching.
We note that a possible straightforward interpretation of lognormal distributions in the mass rate properties (including the BH accretion rates) resides in chaotic cold accretion, which is a turbulence-driven mechanism based on the multiplicative, lognormal process of eddies cascading into progressively smaller eddies \citep{gaspari17c}.

Figures \ref{fig:sfr_combine}--\ref{fig:effhist} reveal several interesting trends, including a tight correlation between the star formation and cooling rate that flattens out at low cooling rates, with a potential upturn at the high-\.M$_{cool}$ end, and with significant scatter in SFR at fixed \.M$_{cool}$. In the following sections we will investigate the SFR--\.M$_{cool}$ relationship in more detail and speculate on the physical origin of each these different features.

\section{The Slope and Scatter of the SFR--\.M$_{cool}$ Relation}

\begin{figure*}[htb]
\centering
\includegraphics[width=0.97\textwidth]{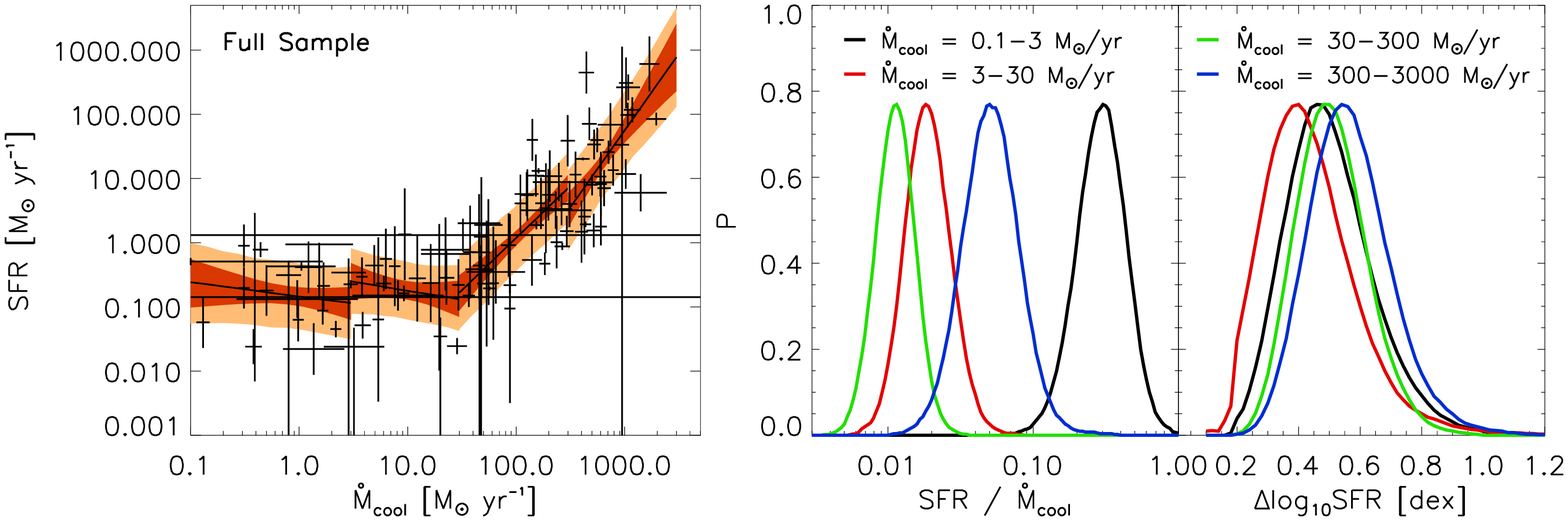}\\
\includegraphics[width=0.97\textwidth]{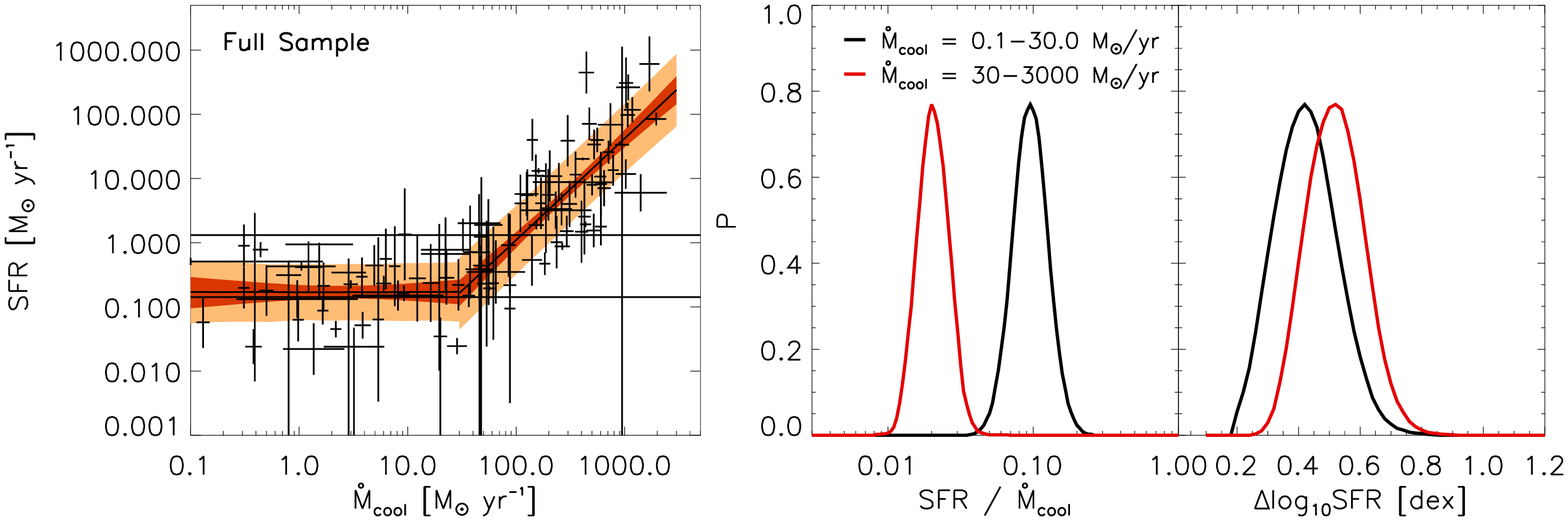}
\caption{\emph{Upper left:} SFR--\.M$_{cool}$ relation, as shown in Figure \ref{fig:sfr_combine}. These data are divided into four chunks in cooling rate, with each chunk being independently fit with a function of the form $SFR = A\dot{M}_{cool}^B \pm C$, where $C$ is a log-normal scatter. The black line shows the best fit, while the dark red region shows the 1$\sigma$ allowable range of fits. The orange region shows the best fit including scatter. \emph{Upper middle:} Best fit values of the cooling efficiency, $\epsilon_{cool} \equiv SFR/\dot{M}_{cool}$. For each fit, we show the allowable value of $\epsilon_{cool}$ at the midpoint of the bin. \emph{Upper right:} Probability distribution for the fit parameter $C$. This panel demonstrates that the scatter is consistent with being constant across all fits. \emph{Lower panels:} The same as above, but considering only two bins in \.M$_{cool}$ rather than four.
}
\label{fig:mcmc_chunks}
\end{figure*}

Figures \ref{fig:sfr_combine} and \ref{fig:sfr_dmdt_averages} suggest that the slope of the \.M$_{cool}$--SFR relation varies as a function of \.M$_{cool}$. To assess this quantitatively, we divide the data into four logarithmically-spaced bins in cooling rate and fit the data in each bin with a function of the form $SFR = A*\dot{M}_{cool}^B \pm C$, where C is a log-normal scatter. These fits were performed using the software {\sc linmix\_err}\footnote{\url{https://idlastro.gsfc.nasa.gov/ftp/pro/math/linmix_err.pro}} \citep{kelly07} which is a Bayesian approach to linear regression that incorporates uncertainties in both parameters, properly treats non-detections, and includes intrinsic scatter in the fitting. For these fits, we include as upper limits those systems for which the AGN contamination could not be fully removed (2) and those for which star formation is not detected (6), yielding 8 upper limits.

We show the results of these fits in Figure \ref{fig:mcmc_chunks}. In the two bins for which \.M$_{cool} < 30$ M$_{\odot}$ yr$^{-1}$, the slope is consistent with flat ($B=-0.22 \pm 0.40$,  $B=-0.25 \pm 0.49$), while at \.M$_{cool} > 30$ M$_{\odot}$ yr$^{-1}$, the slope increases and is consistent with a single value, greater than unity, over two decades in \.M$_{cool}$ ($B=1.62 \pm 0.42$, $B=2.37 \pm 0.74$). For cooling rates of 100 M$_{\odot}$ yr$^{-1}$, a typical value for cool core clusters, we find $\epsilon_{cool} = 0.012 \pm 0.003$. This efficiency appears to be considerably higher for the most massive ($\epsilon_{cool} = 0.05 \pm 0.02$) and least massive ($\epsilon_{cool} = 0.3 \pm 0.1$) systems. There is no evidence for the scatter depending on \.M$_{cool}$, with all four bins having similar scatters of $\sim$0.5 dex.

Given that there is no statistical difference in the slope and scatter between the two low--\.M$_{cool}$ and two high--\.M$_{cool}$ bins, we combine these into wider bins in order to improve the fit statistics. 
The results of these fits are shown in the lower panels of Figure \ref{fig:mcmc_chunks}. We find, for low--\.M$_{cool}$ systems (0.1--30.0 M$_{\odot}$ yr$^{-1}$), that $SFR \propto \dot{M}_{cool}^{0.00\pm0.15}$. This flat slope may be due to a variety of effects, both physical and systematic -- we will address these in detail in \S5. Assuming that the star formation is indeed fueled by the cooling ICM, at the midpoint of this bin (\.M$_{cool} \sim 2$ M$_{\odot}$ yr$^{-1}$) the implied cooling efficiency is $\epsilon_{cool} = 0.10 \pm 0.02$. Due to the slope being flat, this efficiency will continue to rise towards lower values of \.M$_{cool}$.

For high-\.M$_{cool}$ systems (30-3000 M$_{\odot}$ yr$^{-1}$), we measure a slope of $SFR \propto \dot{M}_{cool}^{1.59\pm0.18}$. The fact that this slope is greater than unity implies that more massive systems, which tend to have higher cooling rates, are able to cool more efficiently than their low-mass counterparts. This may be signaling a ``saturation'' of AGN feedback -- we will discuss this further in \S6. The implied cooling efficiency for systems with \.M$_{cool} = 300$ M$_{\odot}$ yr$^{-1}$ is $0.021 \pm 0.004$, which is consistent with the value measured by simply collapsing all of the data into a histogram (Figure \ref{fig:effhist}) and represents the most precise estimate of the cooling efficiency to date in cool core groups and clusters. 

We measure the intrinsic scatter of the SFR--\.M$_{cool}$ relation for systems with \.M$_{cool} > 30$ M$_{\odot}$ yr$^{-1}$, finding a log-normal scatter of $0.52 \pm 0.06$ dex. The similarity between this measurement and the value obtained by simply collapsing all of the data into a histogram (Figure \ref{fig:effhist}) suggests that the large scatter observed in Figure \ref{fig:sfr_combine} is dominated by intrinsic scatter, rather than measurement uncertainties. As discussed in the previous section, this supports a picture in which the cooling--feedback balance is only well-regulated on very long time periods, with short periods of over-cooling and over-heating leading to large scatter in the SFR at fixed \.M$_{cool}$. 

In summary, the median cluster with \.M$_{cool} \sim 300$ M$_{\odot}$ yr$^{-1}$ harbors a BCG in which the SFR is $2.1\pm0.4$\% of the cooling rate, with an intrinsic cluster-to-cluster scatter of $0.52\pm0.06$ dex. This scatter appears to be independent of cooling rate, to the degree with which it can be constrained. The slope of this relation is greater than unity, suggesting that the highest mass systems have more efficient cooling than their low-mass counterparts. At M$_{cool}$ $<$ 30 M$_{\odot}$ yr$^{-1}$, the trend flattens out, with the SFR becoming independent of the cooling rate (SFR $\propto$ \.M$_{cool}^{0.00\pm0.15}$). In the following sections, we will attempt to provide a physical interpretations for each of these various features.

\section{Elevated Star Formation Rates in Slowly-Cooling Systems}

In Figures \ref{fig:sfr_combine}, \ref{fig:sfr_dmdt_averages}, and \ref{fig:mcmc_chunks}, we demonstrate that the SFR in BCGs is constant for cooling rates spanning 0.1 M$_{\odot}$ yr$^{-1}$ to 30 M$_{\odot}$ yr$^{-1}$ (SFR $\propto$ \.M$_{cool}^{0.00\pm0.15}$). For the systems with the lowest \.{M}$_{cool}$, $\epsilon_{cool}$ approaches $\sim$100\%, two orders of magnitude higher than for systems with cooling rates $>$10 M$_{\odot}$ yr$^{-1}$. As discussed in \S3, this trend could be due to one of many scenarios, including (but not limited to) the following: i) the observed star formation in low-mass systems is not due to cooling of the hot ICM, but rather to some other source of cool gas, and so should not be correlated with the cooling rate or have its upper limit bound by the cooling rate; ii) we are missing a large ($N>100$) population of low mass systems with cooling rates of 0.1--1.0 M$_{\odot}$ yr$^{-1}$, and the few systems we do see are $>$3$\sigma$ outliers; or iii) SFRs measured in the lowest-mass systems are biased high, either due to an inability to constrain SFRs as low as $\sim$0.001 M$_{\odot}$ yr$^{-1}$, or a higher fraction of AGN contamination.

\begin{figure}[htb]
\centering
\includegraphics[width=0.49\textwidth]{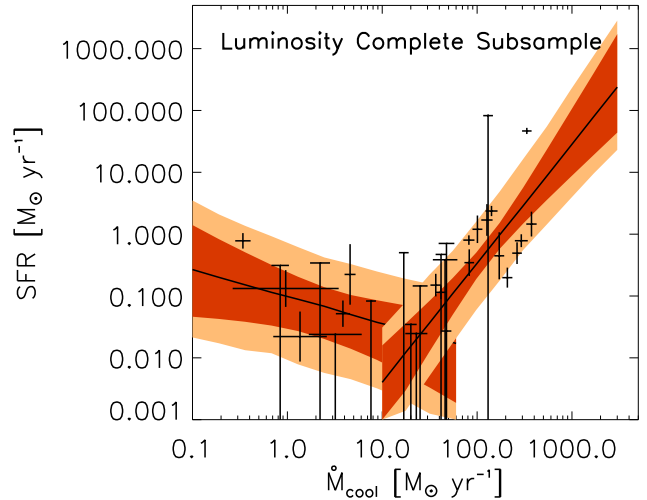}
\caption{Similar to Figure \ref{fig:mcmc_chunks}, but only showing groups and clusters satisfying $L_X > 3.3\times10^{44}$ erg s$^{-1}$ and $z<0.1$ and using uniformly-measured SFRs, including upper limits. We fit a function of the form $SFR = A*\dot{M}_{cool}^B \pm C$ to these systems separately in two regimes, \.M$_{cool} > 10$ M$_{\odot}$ yr$^{-1}$ and \.M$_{cool} < 60$ M$_{\odot}$ yr$^{-1}$, finding consistent fits to the full sample (Figure \ref{fig:mcmc_chunks}). This suggests that the two-slope behavior of the SFR--\.M$_{cool}$ relation is not due to sample incompleteness or biases in the SFR estimates.
}
\label{fig:mcmc_fmchunks}
\end{figure}

We can address the second and third possibilities by considering a luminosity-complete subsample of groups and clusters drawn from \cite{fraser14}, with proper treatment of SFR non-detections. For systems at $z<0.1$ and $L_X > 3.3\times10^{44}$ ergs s$^{-1}$, there is available \emph{Chandra} data for $>$93\% of systems, allowing us to measure \.{M}$_{cool}$ following \S2.2. As discussed in \S2.3, we have remeasured SFRs for each of these systems, carefully applying $k$-corrections and subtracting stellar continuum emission, following \cite{green16}. As a result of this more careful re-analysis, we infer upper limits on the SFR for 13/31 systems, primarily at the low-\.M$_{cool}$ end. Measuring the slope of the SFR--\.M$_{cool}$ relation for systems with $0.1 < \dot{M}_{cool} < 60$ M$_{\odot}$ yr$^{-1}$, and incorporating these non-detections, yields a value of $-0.43_{-0.57}^{+0.35}$, consistent at the $\sim$1$\sigma$ level with the value of $0.00 \pm 0.15$ measured for the full sample. We note that this measurement is made over a slightly larger baseline in \.M$_{cool}$ in order to have enough detections to constrain the slope, scatter, and zero point. At the high-\.M$_{cool}$ end, the slope is poorly constrained ($1.93_{-0.54}^{+0.66}$) due to the lack of massive systems in a volume-limited sample, but it is still consistent with the measurement for the full sample.  These results are shown in Figure \ref{fig:mcmc_fmchunks}, and imply that neither sample completeness or biases towards detections are driving the flattening of the slope at low-\.M$_{cool}$.

\begin{figure}[h!]
\centering
\includegraphics[width=0.49\textwidth]{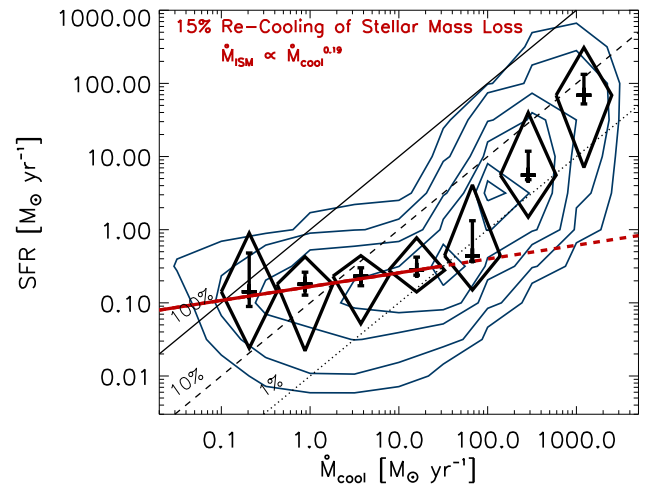}
\caption{Binned averages, as in Figure \ref{fig:sfr_dmdt_averages}, overplotted on the full sample (shown with dark blue contours). We compare the data to a simple toy model, representing the recycling of gas lost by evolved stars (\.M$_{ISM}$; red). Assuming simple scaling relations between the classical cooling rate, the cluster mass, and the BCG stellar mass (see \S5), we expect that \.M$_{ISM}$ $\sim$ 1.07$\times$\.M$_{ICM}^{0.19}$. This model assumes that, given a reservoir of cool gas, star formation is efficient at a level of $\sim$15\% \citep[e.g.,][]{lada03}. This model, which has essentially only one free, but constrained, parameter (re-cooling efficiency) provides an excellent match to the data at the low-\.M$_{cool}$ end, suggesting that the cooling ICM is not providing the fuel for star formation in systems with \.M$_{cool}$ $\lesssim$ 30 M$_{\odot}$ yr$^{-1}$.}
\label{fig:sfr_dmdt_scatter+model}
\end{figure}

Given that the elevated SFRs (compared to the canonical 1\% cooling efficiency) at low values of \.{M}$_{cool}$ are not entirely due to measurement or selection biases, we investigate the potential that the relevant physics is changing as a function of mass. The simplest explanation for the upturn at low \.{M}$_{cool}$ is that there is a secondary source of fuel for star formation in these systems. One possibility is that the frequency of gas rich mergers in groups or isolated ellipticals is higher than in massive, rich clusters. However, looking at the 10 lowest mass systems in our sample, there is no evidence for recent/ongoing merging activity.  Another, more likely, possibility is that mass loss from evolved stars is fueling star formation at a low level in all ellipticals \citep[e.g.][]{mathews03,voit11}. Using total masses from \cite{main17} for 34 overlapping clusters, we can derive an empirical relationship between \.{M}$_{cool}$ and total cluster mass, finding $\dot{M}_{cool} \propto M^{1.8}$. From this, we use the M$_{500}$--M$_{*,BCG}$ relation from \cite{kravtsov14}, and the relation between M$_{*,BCG}$ and the amount of mass lost by evolved stars \citep{mathews03} to derive the amount of available gas from stellar mass loss as a function of ICM cooling rate, finding \.M$_{stars}$ $=$ 1.07$\times$\.M$_{ICM}^{0.19}$.  The slope of this predicted powerlaw relationship is consistent with the value of $0.00\pm0.15$ that we measure for low-mass systems. We show this curve in Figure \ref{fig:sfr_dmdt_scatter+model}, where we have assumed an efficiency of star formation out of the recycled ISM gas of 15\%, consistent with the 10--30\% quoted by \cite{lada03}.

Figure \ref{fig:sfr_dmdt_scatter+model} demonstrates that the flattening of the SFR as a function of \.{M}$_{cool}$ at low values of \.{M}$_{cool}$ can be attributed to the re-cooling of material ejected from evolved stars in the elliptical galaxy, following \cite{mathews03}. Assuming reasonable star formation efficiencies out of this processed material \citep{lada03}, coupled with a consistent $\sim$1\% cooling flow at all masses, allows us to predict the SFR over $\sim$4 orders of magnitude in classical cooling rate. Given that there is no evidence for runaway cooling in these systems -- indeed, the amount of AGN feedback is actually higher for low-mass systems than is needed to offset cooling \citep{rafferty06, nulsen09} -- we find this explanation to be the most plausible.

\section{Elevated Star Formation Rates in the Most Rapidly-Cooling Systems}

\begin{figure*}[htb]
\centering
\includegraphics[width=0.97\textwidth]{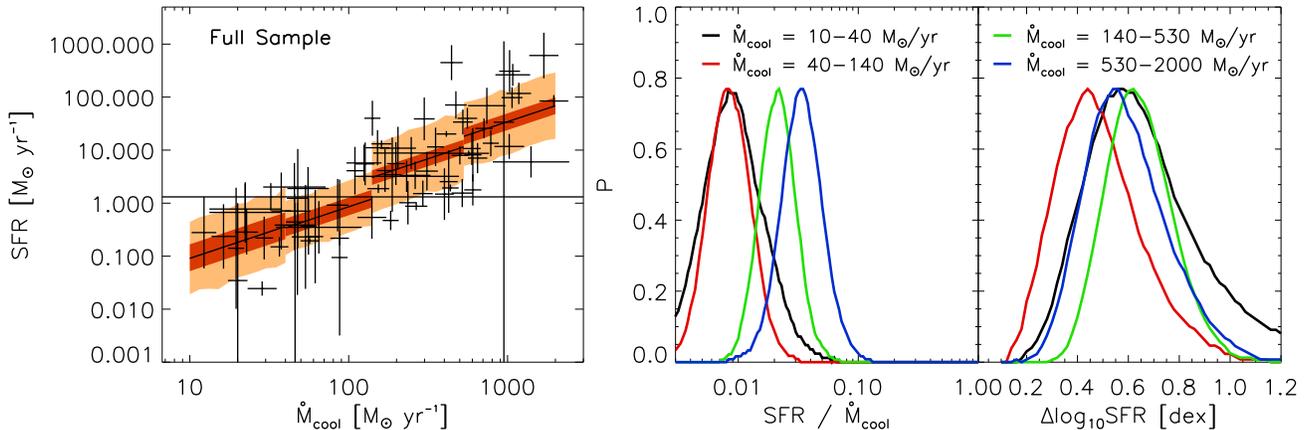}
\caption{Similar to Figure \ref{fig:mcmc_chunks}, but considering only systems with \.M$_{cool}$ $>$ 10 M$_{\odot}$ yr$^{-1}$. We these systems into four chunks in cooling rate. Due to the low number of points per bin, we fix the slope to unity, considering only the change in normalization and scatter as a function of cooling rate. We find no evidence for a \.M$_{cool}$ dependence in the scatter at fixed SFR. On the contrary, we find that the cooling efficiency increases with increasing \.M$_{cool}$, from $\sim$0.8\% at the low-\.M$_{cool}$ end to $\sim$4\% at the high-\.M$_{cool}$ end.}
\label{fig:mcmc_highM}
\end{figure*}

There is some evidence in Figure \ref{fig:mcmc_chunks} for an increase in the slope of the SFR--\.M$_{cool}$ relation at the high-\.M$_{cool}$ end. We investigate this further by splitting the sample of galaxies, groups, and clusters with $10 < \dot{M}_{cool} < 2000$ M$_{\odot}$ yr$^{-1}$ into four bins (10--40, 40--140, 140--530, 530--2000), and fitting each of these individually with a fixed slope of unity. This allows us to determine whether we are seeing a changing slope (which cause a changing normalization over the small bins), a changing scatter, both, or neither. We note that, due to the small number of points per bin, one of the slope, scatter, or normalization \emph{must} be fixed.

In Figure \ref{fig:mcmc_highM} we show the results of this test. For systems in the two bins with \.M$_{cool} < 140$ M$_{\odot}$ yr$^{-1}$, we measure cooling efficiencies of 0.8-1.0\%, consistent with the canonical value of 1\%. Since these two bins have statistically consistent fits, we combine them, finding $\epsilon_{cool} = 0.8_{-0.2}^{+0.3}$\%.
In the $140 < \dot{M}_{cool} < 530$ M$_{\odot}$ yr$^{-1}$ bin, we measure $\epsilon_{cool}=2.2_{-0.6}^{+0.8}$\%, while in the $530 < \dot{M}_{cool} < 2000$ M$_{\odot}$ yr$^{-1}$ bin, we measure $\epsilon_{cool}=3.7_{-1.2}^{+1.7}$\%. These three measurements imply a roughly four-fold increase (at a confidence level of 3.3$\sigma$) in the cooling efficiency of galaxies, groups, and clusters over the range $10 < \dot{M}_{cool} < 2000$ M$_{\odot}$ yr$^{-1}$. We note that, over more than two orders of magnitude in cooling rate, we see no change in the scatter of this relation, with all bins being fully consistent with the value of 0.52 dex measured for the full sample.

The increase in cooling efficiency at the high-\.M$_{cool}$ may be due to selection effects -- we deliberately included systems like Perseus, Phoenix, IRAS09104+4109 and H1821+643 in our sample, all of which have central starburst BCGs. However, in Figure \ref{fig:sfr_combine} we see a clear dearth of systems with cooling efficiencies $>$10\% and cooling rates of 10--300 M$_{\odot}$ yr$^{-1}$. Such systems ought to be well-known, as they would harbor very star-forming BCGs and likely strong AGN feedback. Given that such systems are often the targets of intense multi-wavelength follow-up campaigns, it is unlikely that this dearth of highly-efficient cooling flows is an incompleteness issue. 

\begin{figure}[htb]
\centering
\includegraphics[width=0.49\textwidth]{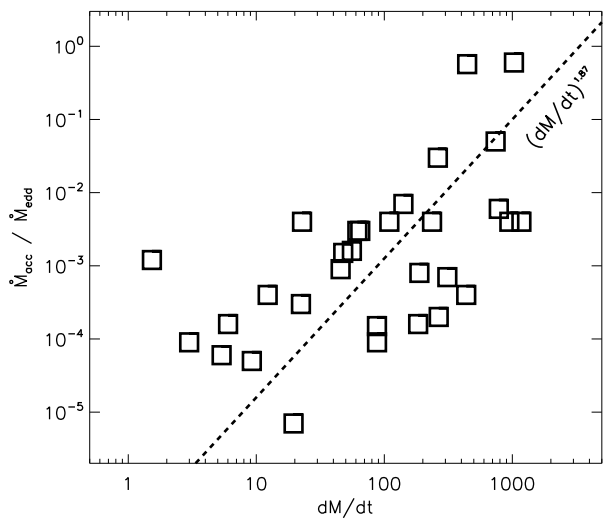}
\caption{Black hole accretion rate, normalized to the Eddington rate, versus ICM cooling rate for 28 galaxies, groups, and clusters from \cite{russell13}. The dashed line represents the expectation assuming empirical scaling relations between the cooling rate, total mass, black hole mass, and accretion rate (see discussion in \S6).
This figure demonstrates that the systems with the highest \.M$_{cool}$ ought to have black hole accretion rates of $>$1\% of Eddington, which would lead to a much higher fraction of radiatively efficient (and thus mechanically inefficient) AGN in the centers of the most massive cool core clusters. This may explain the increased SFR in the most massive systems, as they are more susceptible to ``flickering'' between radiative and mechanical feedback.}
\label{fig:mdotmedd}
\end{figure}

If the star formation rate per unit cooling gas is indeed enhanced in the most massive systems, one of two possibilities emerge: that either the cooling from hot to cold is more efficient or that the conversion of cool gas into stars is more efficient in these systems. To test the latter hypothesis, we would need estimates of the cool gas reservoir for these systems -- we defer such an analysis to a future paper where we will combine these data with new and existing measurements of the gas content for a large sample of galaxies, groups, and clusters. Instead, we will assume here that the increased SFRs indicate an increase in the cooling efficiency of the hot gas in these systems.

In Figure \ref{fig:sfr_combine}, it is clear that the majority of systems hosting strong central AGN also have high cooling rates. The accretion rate onto the central supermassive black hole can be related to the large-scale cooling rate by the following proportionality:

\begin{equation}
\dot{M}_{acc} \propto L_X \propto \dot{M}_{cool}  kT \propto \dot{M}_{cool} M^{0.65} \propto \dot{M}_{cool}^{2.45} ,
\label{eq:macc}
\end{equation}

\noindent{}where we assume that the black hole accretion rate goes like the X-ray luminosity in the core \citep{gaspari17b}, which in turn is proportional to the temperature and cooling rate of the cluster. 
This accretion rate is capped at the Eddington rate, where radiation pressure offsets the gravitational pull of the black hole.  The Eddington rate scales with the black hole mass, which is proportional to the central galaxy mass \citep{mcconnell13}, which is only weakly dependent on the cluster mass \citep{kravtsov14}:

\begin{equation}
\dot{M}_{Edd} \propto M_{BH} \propto M_{BCG}^{1.05} \propto M^{0.32} \propto \dot{M}_{cool}^{0.58} .
\label{eq:medd}
\end{equation}

\noindent{}Combining these two equations, we predict that  \.M$_{acc}$/\.M$_{Edd}$ $\propto$ \.M$_{cool}^{1.87}$. That is, groups and clusters with higher cooling rates ought to have central black hole accretion rates that approach the Eddington rate. 
In Figure \ref{fig:mdotmedd}, we show that this relation provides an adequate description of the data, where black hole accretion rates have been taken from \cite{russell13}. Such a relation can lead to a steepening of the SFR--\.M$_{cool}$ relation in two ways. First, the most massive systems are more likely to reach the Eddington rate, at which point feedback should ``saturate''. This should lead to a flattening of the relation between the cooling rate and the accretion rate and, by extension, the amount of feedback energy.

An alternative explanation is that AGN feedback is less effectively coupled to the ICM at high black hole accretion rates. At high accretion rates, relative to Eddington, the power output from a supermassive black hole transitions from outflow-dominated to radiation-dominated \citep{churazov05}. This transition happens at roughly \.M$_{acc}$/\.M$_{Edd} \sim 0.01$ \citep{russell13}. While radiative feedback can be effective at quenching star formation in galaxies \citep[e.g.,][]{hopkins10}, it may be less effective in the cluster environment. \cite{walker14} showed that H1821+643, which is a radiation-dominated AGN at the center of a massive galaxy cluster, appears to be \emph{cooling} the surrounding hot ICM. The fact that the bulk of the accretion disk is cooler than the $>$10$^7 $K ICM means that the radiation from the AGN will lead to Compton cooling of the hot ICM. This cooling is observed as a rapid decrease in the entropy of the ICM in the inner $\sim$10\,kpc of the cluster. 

Given that the mechanical output of an AGN grows with accretion rate (for low accretion rates), we expect most galaxies, groups, and clusters to be oscillating around a steady state: if a nonlinear condensation develops, the accretion rate will spike, leading to a burst of feedback, which will prevent further cooling. However, for massive clusters where the cooling rate is high, Figure \ref{fig:mdotmedd} tells us that the black hole is accreting at a substantial fraction of the Eddington rate, and will output much of its energy in the radiative mode. The development of a thermal instability may then lead to a burst of \emph{radiative} feedback, which (if H1821+643 is representative of such systems) is unable to quench cooling on large scales. Further, the amount of feedback in such system is naturally capped by the Eddington rate, despite no such cap existing on the large-scale cooling rate. The combination of these effects could lead to the accumulation of massive reservoirs of cold gas in the most massive systems, such as the Phoenix cluster, which can fuel massive starbursts. Such a scenario would lead to more efficient star formation in the most massive clusters (which host the strongest cooling flows), as these systems are the most likely to be accreting near the Eddington rate.

\section{Understanding the Scatter in Star Formation at Fixed Cooling Rate}


We measure a log-normal intrinsic scatter in SFR at fixed cooling rate of $0.52 \pm 0.06$ dex for systems with cooling rates spanning 30--3000 M$_{\odot}$ yr$^{-1}$. This scatter does not appear to vary with cooling rate, nor does it appear to be primarily due to selection effects -- for a luminosity-complete subsample we measure a scatter in the SFR at fixed cooling rate (for \.M$_{cool} > 10$ M$_{\odot}$ yr$^{-1}$) of $0.67_{-0.15}^{+0.22}$ dex, consistent at the 1$\sigma$ level with our measurement for the full sample. This large scatter may be due to a number of different physical processes, including (but not limited to) differing timescales between the SFR and \.M$_{cool}$ measurements, imbalances in the cooling/feedback cycle, or inefficient (i.e., radiative) feedback dominating in some systems. We will discuss each of these points below. 

\begin{figure*}[htb]
\centering
\includegraphics[width=0.98\textwidth]{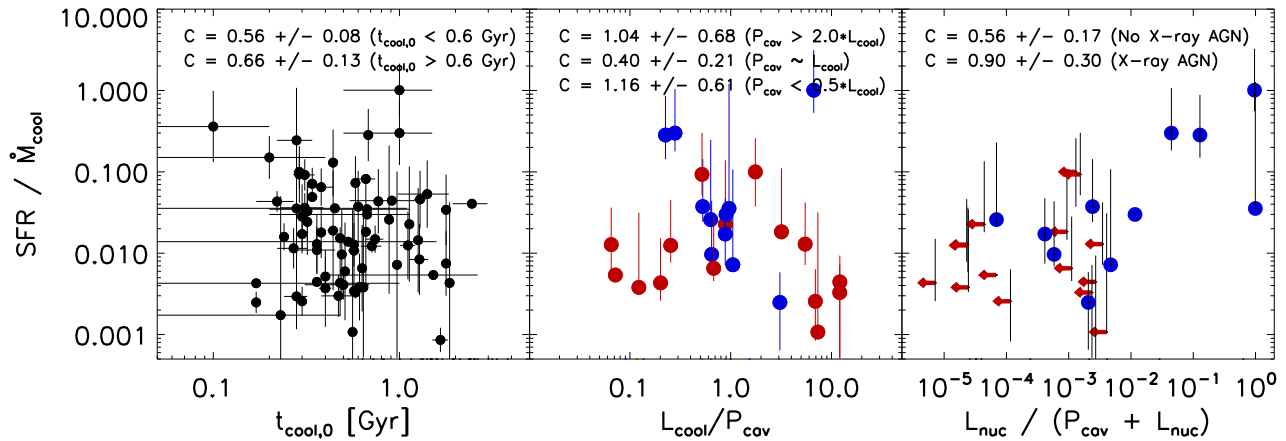}
\caption{Cooling efficiency, $\epsilon_{cool} \equiv SFR/\dot{M}_{cool}$, as a function of the central cooling time (left), the ratio of the cooling luminosity to the jet power (middle), and the fraction of AGN power outputted as radiation (right). In the middle and right panels, blue points represent systems for which an X-ray point source is detected at the cluster center, while red points have upper limits on the X-ray luminosity of the central AGN. This figure demonstrates that the scatter in the SFR--\.M$_{cool}$ relation is uncorrelated with the central cooling time, but does appear to correlate weakly with the properties of the central AGN. In particular, for systems where the cooling is not well-regulated by AGN feedback (i.e., greater than a factor of two difference between the cooling and feedback powers), and systems for which the central AGN is X-ray bright, we measure an increased scatter in $\epsilon_{cool}$.}
\label{fig:scatter}
\end{figure*}

The SFR estimates used here, based on emission lines, UV continuum, and re-radiated dust emission, probe O and B stars, with lifetimes of 1--100 yr. The bulk of this star formation is typically contained within $<$10\,kpc \citep{mcdonald11a,tremblay15}. In contrast, the cooling rates that we calculate are time-averaged on scales of 3\,Gyr, and are measured on scales of $\sim$100\,kpc. This larger aperture is partially motivated by observation (see \S2.2), and partially by necessity -- there is often not sufficient quality X-ray data to quantify the cooling rates on smaller scales. In the left panel of Figure \ref{fig:scatter} we attempt to address this issue, plotting the cooling efficiency as a function of the cooling rate measured in the inner 10 kpc ($t_{cool,0}$). If the dominant source of scatter in the SFR--\.M$_{cool}$ relation is due to mismatching timescales between cooling and star formation, we would expect the scatter to correlate with this more localized measurement. We see no evidence that the scatter is related to the central cooling time, finding consistent scatters between systems with short ($<$0.6 Gyr) central cooling times and long ($>$0.6 Gyr) central cooling times.

Another possibility is that the scatter in the SFR at fixed \.M$_{cool}$ is related to the balance between heating and cooling. We examine this possibility in the middle panel of Figure \ref{fig:scatter}, where we plot the cooling efficiency as a function of the ratio of the cooling luminosity (L$_{cool}$) to the mechanical power of the AGN (P$_{cav}$). For systems where the AGN power is significantly greater than or less than the cooling luminosity, we measure a scatter in SFR at fixed \.M$_{cool}$ of $\gtrsim$1 dex. On the other hand, if we consider systems for which the AGN power is within a factor of two of the cooling luminosity (i.e., well-regulated), the scatter drops to $\sim$0.4 dex. This difference is only marginally significant ($\sim$1$\sigma$). Interestingly, we find a tendency towards low SFR for systems where cooling dominates feedback (the 4 systems with the lowest cooling efficiency are all cooling-dominated systems). This runs counter to the expectation, that cooling ought to proceed more efficiently in systems with under-powered AGN. This may indicate that feedback is needed to ``stimulate'' cooling \citep[following][]{mcnamara16}, leading to enhanced SFRs in systems with stronger feedback, or that there is a significant delay between the disappearance of bubbles and the onset of cooling in the feedback loop.

Finally, we consider the possibility that the \emph{type} of feedback is important in setting the scatter in the SFR--\.M$_{cool}$ relation. In the right panel we show the cooling efficiency as a function of the fraction of the AGN power output in radiation ($L_{nuc}/[P_{cav}+L_{nuc}]$), where L$_{nuc}$ is the radiative power of the nucleus and P$_{cav}$ is the mechanical power. We find here that the scatter is weakly correlated with the fraction of AGN power in the radiative mode, with the most star-forming systems also having the most radiatively efficient AGN. Likewise, if we consider the scatter in $\epsilon_{cool}$ for systems with X-ray-bright AGN versus those without X-ray-bright AGN, we find that the former has a factor of $\sim$2 larger scatter. We note that this trend is likely not driving the scatter over the bulk of the relation, but rather may be driving the upturn in cooling efficiency for the most massive systems, as described in \S6. For systems with AGN accreting at $<$1\% of the Eddington rate, there is no difference in scatter for systems with X-ray bright AGN and without.

There is some evidence that the properties of the AGN are responsible for driving the scatter in cooling efficiency, though with these data we are unable to pin down the exact mechanism responsible for driving the scatter. It appears likely from Figure \ref{fig:scatter} that both radiative feedback and mechanical feedback contribute some amount to the scatter, with minimal scatter being achieved when the mechanical AGN power is well matched to the cooling luminosity. We require a larger sample of systems for which we measure both the SFR and mechanical power of the AGN to say with any certainty if this is the case. We note that there is likely some intrinsic non-zero scatter that is independent of the feedback cycle that can be attributed to the chaotic condensation of cool clouds, which simulations predict ought to have an intrinsic scatter over long periods of time between 0.4--0.8 dex \citep{gaspari12a,gaspari12b}.


\subsection{A Case Study: The Phoenix and RBS797 Clusters}



\begin{deluxetable*}{c c c c c c c c c c c}
\tabletypesize{\footnotesize} 
\tablecolumns{5}
\tablewidth{0pt}
\tablecaption{Properties of the cluster, central BCG, and central AGN for the RBS797 and Phoenix clusters}
\tablehead{
\colhead{Cluster} & 
\colhead{$z$} & 
\colhead{M$_{500}$} & 
\colhead{\.{M}$_{cool}$} & 
\colhead{\.M$_{acc}$/\.M$_{Edd}$} & 
\colhead{P$_{cav}$} & 
\colhead{$t_{buoy}$} &  
\colhead{K$_{0}$} & 
\colhead{$t_{cool,0}$} &  
\colhead{L$_{X,AGN}$} & 
\colhead{SFR} \\
 & & 
 \colhead{[$10^{15}$ M$_{\odot}$]} & 
 \colhead{[M$_{\odot}$/yr]} & & 
 \colhead{[$10^{45}$ erg/s]} & 
 \colhead{[$10^7$ yr]} & 
 \colhead{[keV cm$^2$]} & 
 \colhead{[Gyr]} & 
 \colhead{[$10^{44}$ erg/s]} & 
 \colhead{[M$_{\odot}$/yr]}}
\startdata
RBS797 & 0.354$^a$ & 1.2$^a$ & 1404 & 0.025 & 3--6$^a$ & 2--4$^a$& $\sim$20$^{a*}$ & 0.2 &  2$^a$ & 1--10$^a$\\
Phoenix & 0.597$^b$ & 1.3$^c$ & 1691 & 0.050 & 2--7$^b$ & 2--6$^b$ &19$^b$ & 0.3  & 56$^b$ & 610$^b$
\enddata
\tablecomments{These two clusters are remarkably similar in cluster mass, cooling rate, AGN power output, bubble age, and core ICM properties. The only obvious differences between these two systems are the bolometric AGN luminosities and BCG star formation rates.
References are: $^a$: \cite{cavagnolo11}; $^b$: \cite{mcdonald15b}; $^c$: \cite{mcdonald12c}. Values marked with an asterisk have been estimated by eye from figures in \cite{cavagnolo11}.
}
\label{table:rbs797}
\end{deluxetable*}

An interesting pair of systems to examine in detail are the Phoenix and RBS797 clusters, which provide a unique view of the scatter in SFR at fixed cooling rate. Table \ref{table:rbs797} provides a comparison of these systems, showing that their total mass, cooling rate, black hole accretion rate, AGN mechanical power, central entropy, and central cooling time are all remarkably similar. These both appear to be systems with strong cooling (\.{M}$_{cool}$ $>$ 1000 M$_{\odot}$ yr$^{-1}$) that is being offset by a recent ($t_{buoy} \sim 30$ Myr) outburst of powerful (P$_{cav}$ $\sim 5\times10^{45}$ erg/s) AGN activity. The only major differences between these two systems are their nuclear X-ray luminosity, which differ by a factor of $\sim$25, and their BCG star formation rates, which differ by a factor of $\sim$100. These two systems span the full range of SFR observed in the most massive, strongly-cooling systems (see Figure \ref{fig:sfr_combine}). 

It may be that we are observing two very similar systems at different epochs in the heating/cooling cycle. The star formation rate in RBS797 was derived based on a UV luminosity, which is $\sim$25 times fainter than for Phoenix. If star formation was quenched a short time ago in RBS797, we would expect much fainter UV emission for a fixed starburst mass. As an example, a starburst quenched 60 Myr ago would have an order of magnitude less near-UV flux than an ongoing starburst \emph{for the same total mass formed}. In this way, RBS797 could have had the same SFR as Phoenix, but we are seeing it shortly after quenching. This scenario is unlikely for two reasons. First, the timescales needed to reconcile the different UV fluxes from these two systems is $>$100 Myr, which is substantially longer than the buoyant rise time of the bubbles ($\sim$30 Myr). Second, there is evidence for strong H$\beta$ emission in RBS797, which is indicative of a population of massive young stars, suggesting ongoing star formation. Thus, we conclude that there are, in fact, vastly different amounts of stars being formed in these two systems.


The black hole accretion rate of Phoenix is double that of RBS797, yet the radio (mechanical) output is similar. The energy released by this additional accretion appears to be purely radiative, with Phoenix having an equal split between radiative and mechanical power output \citep[$P_{rad}/(P_{rad}+P_{mech}) \sim 60$\%;][]{mcdonald15b}. For contrast, the power output of RBS797 is dominated by mechanical feedback, with only $\sim$4\% of the power in radiation. This may be the reason for the huge difference in SFR between these two systems. If all of the feedback energy in Phoenix were, instead, purely mechanical, it would be able to quench an additional $\sim$1000 M$_{\odot}$ yr$^{-1}$ of cooling (assuming $\frac{dM}{dt} = \frac{2L_{cool}\mu m}{5kT}$), potentially halting the massive starburst. Thus, it may be that the AGN in Phoenix recently switched to a mix of radiative and mechanical feedback, which has opened the door for a short burst of runaway cooling. At lower cluster masses (and lower cooling rates), the typical black hole accretion rate is much lower (see Figure \ref{fig:mdotmedd}), which means that chaotic, order of magnitude, variations in the accretion rate will never lead to near-Eddington accretion, while in the highest mass (and most rapidly cooling) clusters, it requires relatively small fluctuations in accretion rate to approach near-Eddington. Assuming chaotic cold accretion \citep{gaspari15a,tremblay16}, we expect high-mass systems to go through Phoenix-like and RBS7-like phases more often than low-mass systems, due to the fact that the accretion rate is oscillating around 10$^{-2}$ \.M$_{edd}$ rather than 10$^{-4}$ \.M$_{edd}$.

It is also worth noting that the jets in RBS797 appear to have recently precessed by 90 degrees over the course of multiple AGN outbursts \citep{doria12}. This would lead to a more isotropic heating than in a system with non-precessing jets, and may explain why this system appears to have minimal cooling. Identifying a large sample of clusters for which we observe precessing radio jets would allow us to quantify whether these systems are more effective at quenching cooling of the hot ICM.

The structure of the bubbles (size, distance from center, buoyant rise time) are similar between these two systems, with the only obvious difference being that Phoenix has a greater fraction of its energy output in the radiative mode, while the jets in RBS797 appears to be rapidly precessing. The net result of these difference is to lower the heating rate in Phoenix and raise the isotropic heating rate in RBS797. 
As shown in \cite{gaspari17c}, due to the recurrent inelastic collisions, chaotic cold accretion drives very rapid variability with a flicker noise power spectrum ($-1$ slope in frequency space). This means that we expect a few percent of times a few orders magnitude variations in the SMBH accretion rate. Such large variations will produce a near Eddington event and a temporary transition to the less efficient radiative feedback mode.
This may be what is happening in Phoenix, and not happening in RBS797. These two systems, along with all of the rapidly-cooling systems presented in Figure \ref{fig:sfr_combine}, paint a picture in which short-term cooling can span roughly 3 orders of magnitude, while long-term cooling is well regulated.


\section{Redshift Evolution of the Cooling Flow Problem}

One might expect there to be a redshift dependence to the mean value and scatter in $\epsilon_{cool}$ if it takes some time for the cooling/feedback cycle to ``settle'' into its present state. Recent studies by  \cite{webb15}, \cite{mcdonald16}, and \cite{bonaventura17} have shown that the BCG SFR increases by a factor of $\sim$100 between $z\sim0$ and $z\sim1.5$ for rich clusters of galaxies. Over the same timeframe \cite{mcdonald13b} showed that there was little evolution in \.{M}$_{cool}$ for massive clusters. Taken together, these results would suggest a strong evolution in the cooling efficiency of the ICM over the past $\sim$10 Gyr. However, the star-forming BCGs at high-$z$ are found predominantly in unrelaxed clusters, suggesting that the origin of the SFR may come from galaxy-galaxy interactions, rather than ICM cooling \citep{mcdonald16}. It remains unclear how the BCG SFR evolves in relaxed systems, and how that evolution is dependent on the cooling properties of the host cluster.

\begin{figure}[htb]
\centering
\includegraphics[width=0.49\textwidth]{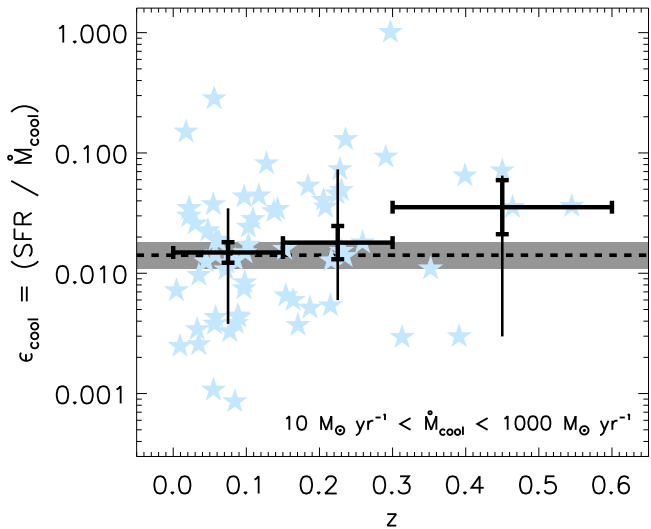}
\caption{Cooling efficiency, $\epsilon_{cool}$ as a function of redshift. Individual systems are shown in light blue, while binned averages are shown in black. Thick error bars represent the uncertainty on the mean, while the thin bars represent the measured 1$\sigma$ scatter. This plot shows that there is no statistically significant evolution in the cooling efficiency over the narrow redshift range probed here.}
\label{fig:sfr_dmdt_redshift}
\end{figure}

In Figure \ref{fig:sfr_dmdt_redshift} we investigate the evolution of $\epsilon_{cool}$ over the redshift range $0 < z \lesssim 0.5$. We find no statistically significant evolution in the ratio of the BCG SFR to the ICM cooling rate over this redshift range. There is a very weak trend towards higher values of $\epsilon_{cool}$ at higher redshifts, but we note that this sample is biased towards high-mass systems at high-$z$, while containing a mostly representative mass distribution at low-$z$. Given the higher cooling efficiencies in the highest \.{M}$_{cool}$ systems, one might expect to observe an artificial increase in $\epsilon_{cool}$ in this plot due to selection effects. Given the small number of systems at $z\gtrsim0.3$ in this sample, we have little ability to probe the redshift evolution of $\epsilon_{cool}$ with this sample, and defer a proper measurement to a follow-up study focusing on a well-defined sample of high-$z$ clusters (McDonald et al.\ in prep).

\section{Summary}

We have assembled a large, inhomogeneous sample of 107 galaxies, groups, and clusters spanning $\sim$3 orders of magnitude in mass, $\sim$4 orders of magnitude in ICM cooling rate, $\sim$5 orders of magnitude in BCG SFR, and $\sim$5 orders of magnitude in black hole accretion rate. For each system, we measure the ICM cooling rate, \.{M}$_{cool}$, using available \emph{Chandra} data, and obtain the BCG SFR and an estimate of the systematic uncertainty in this quantity by carefully combining over 330 SFR estimates in the literature. With these data, we consider how the BCG SFR correlates with the cooling rate of the ICM, finding:

\begin{itemize}

\item For systems with \.{M}$_{cool} > 30$ M$_{\odot}$ yr$^{-1}$, we find that the cooling efficiency ($\epsilon_{cool} \equiv SFR/\dot{M}_{cool}$) is distributed log-normally, with a peak value of $\epsilon_{cool} = 1.4 \pm 0.4$\% (Figure \ref{fig:effhist}) and an intrinsic scatter of $0.52 \pm 0.06$ dex (Figure \ref{fig:mcmc_chunks}). This large scatter implies that the cooling-feedback cycle is only well balanced over long time periods, with BCGs having orders of magnitude different SFR for fixed \.{M}$_{cool}$.

\item For systems with \.{M}$_{cool} > 30$ M$_{\odot}$ yr$^{-1}$, we measure a slope in the SFR--\.M$_{cool}$ relation of $1.59\pm0.18$, suggesting that cooling is more efficient in the highest mass systems (Figure \ref{fig:mcmc_chunks}). Specifically, we find that systems with $\dot{M}_{cool} < 140$ M$_{\odot}$ yr$^{-1}$ have median cooling efficiencies of $0.8_{-0.2}^{+0.3}$\%, while those with $530 < \dot{M}_{cool} < 2000$ M$_{\odot}$ yr$^{-1}$ have median cooling efficiencies of $3.7^{1.7}_{1.2}$\%, nearly a factor of five increase. We propose that this may be due to more rapidly-cooling clusters hosting central black holes accreting at a higher fraction of the Eddington rate (Figure \ref{fig:mdotmedd}), leading to potential saturation of the feedback energy, and/or a transition in the dominant feedback mode from mechanical to radiative at high accretion rates.

\item For systems with \.{M}$_{cool} < 30$ M$_{\odot}$ yr$^{-1}$, we measure a weakening of the correlation between SFR and \.{M}$_{cool}$, such that SFR $\propto$ \.M$_{cool}^{0.00\pm0.15}$ (Figure \ref{fig:mcmc_chunks}). We show that this is not due to a selection effect or due to neglecting upper limits on SFR for non-star-forming systems (Figure \ref{fig:mcmc_fmchunks}). This flat slope is fully consistent with predictions for the re-cooling of stellar mass loss in AGB stars in the central galaxy (Figure \ref{fig:sfr_dmdt_scatter+model}). This implies that, in the average system with \.{M}$_{cool} < 30$ M$_{\odot}$ yr$^{-1}$, star formation is not linked to residual cooling of the ICM.

\item The scatter in $\epsilon_{cool}$ appears to be constant with \.{M}$_{cool}$. We see no evidence that the scatter is related to the cooling timescales (i.e., systems with very different central cooling times exhibit similar scatter). We find weak evidence that the scatter is related to the properties of feedback, with mildly increased scatter for systems with X-ray bright AGN and with AGN powers significantly different than the cooling luminosity. For systems with well-regulated cooling, the scatter is reduced to $\sim$0.4 dex.

\item We present a comparison study, between the RBS797 and Phoenix clusters. These systems have remarkably similar properties in terms of their ICM (total mass, cooling rate, central entropy, central cooling time) and their radio AGN (jet power, bubble ages). Where they differ is a factor of $\sim$100 difference in SFR, a factor of $\sim$25 in AGN luminosity, and the fact that RBS797 appears to have precessing jets. We propose that these systems may represent extrema in the cooling/heating cycle, where occasional short-lived spikes in accretion can lead to radiatively-efficient feedback, preventing the efficient heating of the ICM for a short period of time, while precessing jets can lead to more isotropic heating, leading to maximally suppressed cooling.

\end{itemize}

This study presents firm constraints on the slope and scatter of the SFR--\.M$_{cool}$ relation for low-redshift galaxies, groups, and clusters. It remains an open problem how this relation evolves, or what physical processes drive the large scatter in SFR for fixed \.{M}$_{cool}$. We intend to address these two questions via follow-up studies.


\section*{Acknowledgements} 
We would like to thank Helen Russell for reading multiple drafts of the paper and providing valuable comments.
M.\ M.\ acknowledges support by NASA through contracts G06-17112A and HST GO-14352.
M.\ G.\ is supported by NASA through Einstein Postdoctoral Fellowship Award Number PF5-160137 issued by the Chandra X-ray Observatory Center, which is operated by the SAO for and on behalf of NASA under contract NAS8-03060. Support for this work was also provided by Chandra grant GO7-18121X.
B.\ R.\ M.\ acknowledges generous support from NSERC and the Canadian Space Agency.


%
%

\end{document}